\documentclass[PRD,showpacs,preprintnumbers,amsmath,amssymb]{revtex4}

\usepackage{graphicx}
\usepackage{dcolumn}
\usepackage{bm}

\begin{document}


\title{Gravitational-Wave Signature of an Inspiral into a
Supermassive Horizonless Object}

\author{Michael Kesden, Jonathan Gair, and Marc Kamionkowski}
\affiliation{California Institute of Technology, Mail Code 130-33,
Pasadena, CA 91125}

\date{\today}

\begin{abstract}
Event horizons are among the most intriguing of general
relativity's predictions.  Although on firm theoretical footing,
direct indications of their existence have yet to be observed.  With
this motivation in mind, we explore here the possibility of
finding a signature for event horizons in the gravitational
waves (GWs) produced during the inspiral of stellar-mass compact objects (COs)
into the supermassive ($\sim 10^6 M_\odot$) objects that lie at the center
of most galaxies.  Such inspirals will be a major source for LISA,
the future space-based GW observatory.  We contrast supermassive black
holes with models in which the central object is a supermassive
boson star (SMBS).  Provided the COs interact only
gravitationally with the SMBS, stable orbits exist not just
outside the Schwarzschild radius but also inside the surface of
the SMBS as well.  The absence of an event horizon allows GWs
from these orbits to be observed.
Here we solve for the metric in the interior of a
fairly generic class of SMBS and evolve the trajectory of an
inspiraling CO from the Schwarzschild exterior through the
plunge into the exotic SMBS interior.  We calculate the approximate waveforms
for GWs emitted during this inspiral.  Geodesics within the SMBS
surface will exhibit extreme pericenter precession and other
features making the emitted GWs readily distinguishable from
those emitted during an inspiral into a black hole.
\end{abstract}

\maketitle

\section{Introduction} \label{S:intro}

The black-hole event horizon, a surface from beyond
which no information can be received, is one of the most
intriguing predictions of general relativity.  From the
theoretical point of view, the prediction is on a fairly firm
footing.  However, event horizons have become so central to
physics and astrophysics that efforts to obtain direct empirical
evidence for their existence are certainly warranted.  In other
words, is there any observation or measurement that we can make
that would allow us to ``see'' the event horizon?

This question has motivated a small body
of work in which the phenomenology of accretion onto
alternatives to stellar-mass or supermassive black holes (SMBHs)
has been worked out \cite{torres,XRB}.  X-ray spectroscopy or
$\mu$-arcsec imaging 
might distinguish the accretion disks of black holes from those
surrounding more exotic alternatives \cite{RSline}. A second
approach involves the 
``shadow'' cast on background sources by the black hole when acting as a strong
gravitational lens \cite{lense}.  Current observations are
within a factor of two
of being able to resolve the shadow of Sgr A$^\ast$
using very long baseline interferometry (VLBI) in the submillimeter.
Although these calculations may not be applicable if the predictions
of general relativity hold true on astrophysical scales,
such work can be interesting in its own
right and sometimes illuminates certain aspects of the
phenomenology of the standard black-hole spacetime.

With these motivations in mind, we study here a third possibility, namely whether
gravitational-wave (GW) signals from the inspiral of stellar-mass compact
objects (COs) into the supermassive objects at galactic centers may ultimately
be used to ascertain the existence of event horizons.
There are in fact good prospects of detecting GWs from such extreme-mass-ratio
inspirals (EMRIs).
Measurements of stellar velocities within galactic cusps imply
that the dynamics within a few parsecs of the center are dominated
by a supermassive object \cite{richstone,kormendy}.
Stellar clusters sufficiently dense to enclose enough mass within the
observationally determined volume would have lifetimes much less than the age
of the Galaxy \cite{maoz} leading most to believe that the
object must be compact; i.e., a SMBH.
Still, stellar-dynamical measurements probe distances much
larger than the Schwarzschild radius, and thus cannot distinguish SMBHs from
sufficiently compact alternatives.
Although future stellar-dynamical observations will probe
smaller radii, they will never probe the spacetime structure
anywhere even close to the horizon \cite{nevin}.

These supermassive compact objects are expected to
capture stars from the surrounding
galactic cusps \cite{SigRees,Sig}.  Main-sequence stars will generally be
disrupted at a tidal radius $r_t \approx 2 (M/m_\ast)^{1/3} r_\ast$, where
$m_\ast$ and $r_\ast$ are the stellar mass and radius and $M$
the mass of the supermassive central object.  For stars with a solar
mass and radius, this implies $r_t \approx 50 R_s M_{6}^{-2/3}$ with $M_6$ the
mass of the central object in $10^6 M_\odot$.  Thus main-sequence stars are
tidally disrupted long before they reach the Schwarzschild
radius $R_s=2GM/c^2$.  However an
evolved stellar population will also contain a fraction of white dwarfs,
neutron stars, and stellar-mass black holes that, owing to their smaller radii,
can maintain their integrity down to the innermost stable orbit (ISO) and
beyond.  Under the extreme--mass-ratio approximation,
well justified for the case $\eta \equiv m_\ast/M \lesssim 10^{-6}$, these
COs will travel along geodesics of the central object.
However the galactic cusp is a very crowded environment, and two-body
scattering will change the orbital parameters of the COs over a relaxation time
$t_r$.  Dynamical friction will cause heavy objects such as neutron stars
and stellar-mass black holes to sink to the center of the cusp.
Eventually, the COs will enter the critical region of phase space known as the
loss cone in which the timescale for the loss of energy due to
gravitational radiation is less than the timescale for two-body scattering out
of the loss cone.  After this point,  the evolution of the orbit should be
entirely determined by the loss of energy and angular momentum in GWs,
ultimately leading to capture by the central object.  The expected rate of
such captures is highly model-dependent, subject to uncertainties in the
galaxy luminosity function, the mass function of central objects, and the
initial stellar-mass function (IMF) among other variables.  Conservatively,
event rates of order $10^{-8}$ per galaxy per year are anticipated, implying
0.1 captures per year out to 1 Gpc \cite{SigRees}.  This result could be
enhanced by an order of magnitude by a top-heavy IMF, either due to low
metallicity in the early universe or starbursts in the high-density
environment of the galactic cusp itself \cite{Sig}.  EMRIs about a
$10^6 M_\odot$ central object would produce GWs in the frequency band
$10^{-4}$ to $10^{-2}$ Hz probed by LISA, making them an
interesting subject for theoretical investigation.  

To perform this investigation, we need to contrast a
Schwarzschild black hole with a specific candidate whose
spacetime is identical to Schwarzschild at large distances but
merges smoothly onto a horizonless solution in an interior
region.  For this purpose, we adopt as a ``straw man''
the spacetime of a supermassive boson star (SMBS) whose radius
is only a few times the Schwarzschild radius.  Such a star
consists of a coherent scalar-field configuration
supported against gravity by 
its own self-interaction.  Although
no fundamental scalar fields have yet been discovered, the
fertile imaginations of particle theorists have provided no
shortage of candidates; e.g., the standard-model Higgs
field, the squark, slepton, and sneutrino fields in supersymmetric
models, the axion field, and the dilaton in supergravity models.
We restrict our attention to (nontopological) soliton stars,
which are characterized by interaction potentials for which
bound, stable solutions exist even in the absence of gravity
\cite{BSRev}.  This model allows us to choose parameters for the
scalar-field Lagrangian so that a $\sim 10^6 M_\odot$ SMBS
emerges for massive parameters of order 100 GeV.  This is
not to suggest however that such a SMBS is necessarily likely to arise
within the standard model or its most natural extensions.  SMBSs
with similar structure can just as easily arise with vastly
different mass scales in other models, including
``mini-boson stars'' \cite{miniBS}, nonspherical (but axially symmetric)
scalar-field configurations partially supported against gravitational collapse by
angular momentum \cite{ryanBS}, and
(non-solitonic) boson stars in which a massive scalar
field is held up against gravitational collapse by a quartic
self-interaction \cite{colpi}.

How might the EMRI into such a SMBS differ from that into a SMBH?
The famous ``no hair'' theorem states that the properties of an uncharged
black hole are uniquely determined by the hole's mass $M$ and spin $a$.  Any
stationary, axisymmetric metric can be expanded in terms of mass
and current multipole moments $M_l$ and $S_l$ \cite{hansen}.  For the Kerr
metric, the ``no hair'' theorem implies that all multipole moments can be
expressed in terms of the mass and spin, $M_l + i S_l = M(ia)^l$.  By contrast,
no such strict relation need exist between the multipole moments of a generic
boson-star metric.  For example, Ryan \cite{ryanMM} showed that
all the multipole moments
of the central object can be extracted from the gravitational waveform
produced during an EMRI, even in the restricted case of circular orbits in the
equatorial plane.  As the orbital frequency $\Omega$ increases during
inspiral, the number of radians of orbital motion per logarithmic frequency
interval $d\Phi/d(\ln \Omega)$ can be calculated in a power series in 
$\Omega$.  The coefficients of this power series are simple polynomial
functions of the multipole moments.  Ryan applied this formalism to spinning
boson stars whose mass quadrupole moment $|M_2|$ greatly exceeded $Ma^2$, the
value expected for a black hole of comparable mass and spin \cite{ryanBS}.
The spherical boson stars we consider here have perfectly Schwarzschild
spacetimes outside of their surfaces, making them indistinguishable from black
holes during the early stages of an EMRI.  We rely instead on the GWs produced
following the final plunge from the ISO into the central object itself.  For a
black hole, the presence of an event horizon precludes all observations of the
inspiraling CO subsequent to the final plunge.  After a brief ``ringdown''
period, the black hole ceases to be a significant source of gravitational
radiation.  For boson-star inspirals however, many orbits within the boson-star
interior are expected provided that the CO interacts only gravitationally with
the scalar field.  For compact boson stars with surfaces interior to the ISO,
circular orbits can develop an extremely large post-plunge eccentricity,
leading to a sudden excitation of higher-order harmonics of the fundamental
frequency $f \equiv \Omega/\pi$.  As the CO spirals deeper into
the boson-star potential well, we expect the fundamental
frequency to decrease as less mass is enclosed by smaller
orbits.  By explicitly comparing the waveforms produced during
black-hole and boson-star inspirals, we hope to determine how
effectively they can be differentiated.

A final caveat to consider is whether the accumulation of a large mass at the
center of the boson star either through inspirals or accretion will cause it to
collapse into a black hole.  While we expect such a collapse to occur beyond a
certain mass limit, the calculation of this collapse is highly model-dependent and
beyond the scope of this paper.  Such collapses may affect the event rates of
boson-star inspirals, but will not alter either the dymanics or waveforms which
are the subject of this paper.

The paper is organized as follows.  In \S \ref{S:BS}, we describe the
particular boson-star model examined in this paper as originally formulated by
Ref. \cite{BS}.  Spherically symmetric solutions for the metric
and scalar field
are obtained from the Euler-Lagrange and Einstein equations.  In \S
\ref{S:geo}, we determine the geodesics of the boson-star metric, and consider
qualitatively the possible allowed trajectories for the CO during inspiral.
We then quantify this approach in \S \ref{S:GR} by presenting a model for the
loss of energy and angular momentum to gravitational radiation.
The CO's orbital parameters are then evolved in light of these radiative losses
in \S \ref{S:OE}.  Calculated trajectories and waveforms for
several initial conditions are displayed in \S \ref{S:res}, and some final
remarks on the limitations of our approach and open questions that need to be
addressed are given in \S \ref{S:disc}.  A brief appendix examines the accuracy of
the quadrupole approximation near the ISO.

\section{Boson Star Model} \label{S:BS}
Two requirements must be satisfied for a scalar field to have
nontopological-soliton solutions \cite{BSRev}.  The first of
these requirements is the
existence of an additive conservation law, which by Noether's theorem can be
guaranteed by a symmetry of the Lagrangian.  In the model of Friedberg, Lee, 
and Pang \cite{BS} adopted in this paper, the Lagangian density $\mathfrak{L}$
is invariant under a global phase transformation $\phi \to e^{i\theta} \phi$ of
the complex scalar field $\phi$,
\begin{equation} \label{E:Lag}
\mathfrak{L} = - \phi^{\dagger \mu} \phi_{\mu} - U(\phi^{\dagger} \phi) \, ,
\end{equation}
where a dagger denotes Hermitian conjugation and $\phi_\mu \equiv \partial
\phi / \partial x^\mu$.  Such a Lagrangian will possess a conserved
Noether current
\begin{equation} \label{E:curr}
j^{\mu} \equiv -i(\phi^{\dagger} \phi^{\mu} - \phi^{\dagger \mu} \phi) \, ,
\end{equation}
and a corresponding conserved global charge or particle number
\begin{equation}
N \equiv \int \, j^0 |g|^{1/2} dx^1 dx^2 dx^3 \, ,
\end{equation}
where $|g|$ is the absolute value of the determinant of the metric
$g_{\mu\nu}$.  A nonzero charge $N$ implies a time-varying scalar field by the
definition of the current in Eq.~(\ref{E:curr}); this time dependence is given
by
\begin{equation} \label{E:TD}
\phi(r,t) = \frac{1}{\sqrt{2}} \sigma(r) e^{-i\omega t} \, ,
\end{equation}
where the frequency $\omega$ is independent of position but varies with
particle number $N$.  The second requirement for the existence of soliton
solutions is a constraint on the interaction potential $U(\sigma^2)$.  For
some forms of this potential free particles or a black hole will be
energetically favored over a boson star for all values of $N$.  However if
$U(\sigma^2) - 1/2 m^2 \sigma^2$ is negative for some range of $\sigma$, then
stable boson-star solutions are guaranteed to exist for some values of $N$.
The simplest polynomial form of the potential that satisfies this criterion is
\begin{equation} \label{E:pot}
U = \frac{1}{2} m^2 \sigma^2 \left[ 1 - \left( \frac{\sigma}{\sigma_0}
\right)^2 \right]^2 \, ,
\end{equation}
which allows for a false vacuum within the boson star for $\sigma = \sigma_0$.

\begin{figure}[t!]
\scalebox{.50}{\includegraphics{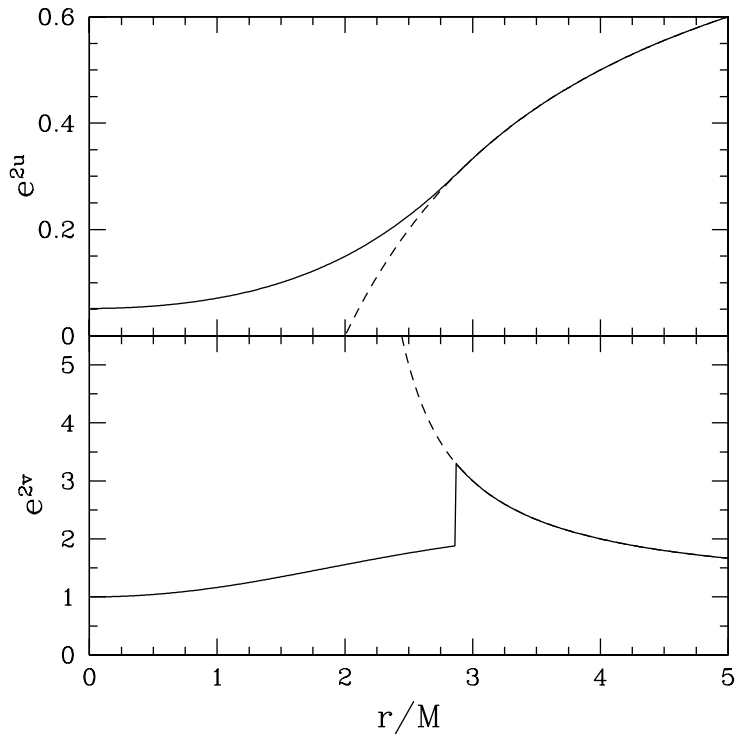}}
\caption{The metric functions $e^{2u(r)}$ and $e^{2\bar{v}(r)}$ as functions
of radius.  The solid curves correspond to a boson star with $R = 2.869 M$
while the the dashed curves are for a black hole of the same mass.  Outside
the boson-star surface the two curves are identical, while for $r < R$ the
solid curves are a numerical solution to Eq.~(\ref{E:metODE}) and the
dashed curves are the Schwarzschild metric functions
$e^{2u(r)} = (1 - 2GM/r)$, $e^{2\bar{v}(r)} = (1 - 2GM/r)^{-1}$.
\label{F:met}}
\end{figure}

Having specified the form of the Lagrangian density $\mathfrak{L}$, we can now
determine the allowed boson-star solutions \cite{BS}.  In the case of
spherical symmetry, the spacetime metric of the boson star can be given in
full generality by
\begin{equation} \label{E:metric}
ds^2 = -e^{2u(r)} dt^2 + e^{2\bar{v}(r)} dr^2 + r^2 (d\theta^2 + \sin^2 \theta
d\phi^2) \, .
\end{equation}
The two metric functions $u(r)$ and $\bar{v}(r)$, along with the 
scalar-field configuration $\sigma(r)$ fully specify a
particular boson-star solution.  They
are chosen to satisfy three independent, ordinary differential equations: the
Euler-Lagrange equation for the scalar field, and the $tt$ and $rr$ components
of the Einstein equation.  In general these equations are stiff; the scalar
field drops from $\sigma \simeq \sigma_0$ to zero across a surface layer of
thickness $m^{-1}$, while the metric functions vary over the much longer
length scale $R \simeq GM$.  If
$\lambda \equiv \sigma_0/m_{\rm Pl} \ll 1$, the scalar field can be described
by a step function
\begin{eqnarray} \label{E:SF}
\sigma &=& \sigma_0 \quad \quad r \leq R \nonumber \\
&=& 0 \quad \quad \, \, \, r > R \, .
\end{eqnarray}
The system of equations then reduces to {\it two} coupled, first-order
equations for the metric functions interior to the boson-star surface at
$r = R$.  With the definitions
\begin{equation} \label{E:bardef}
\bar{r} \equiv \lambda^2 mr \quad {\rm and} \quad e^{-\bar{u}} \equiv
\frac{\omega}{\lambda m} e^{-u} \, ,
\end{equation}
these two equations are
\begin{eqnarray} \label{E:metODE}
2 \bar{r} \frac{d\bar{v}}{d\bar{r}} &=& \left( \frac{1}{2} e^{-2\bar{u}}
\bar{r}^2 - 1 \right) e^{2\bar{v}} + 1 \, , \nonumber \\
2 \bar{r} \frac{d\bar{u}}{d\bar{r}} &=& \left( \frac{1}{2} e^{-2\bar{u}}
\bar{r}^2 + 1 \right) e^{2\bar{v}} - 1 \, .
\end{eqnarray}
Physical boundary conditions imply that $\bar{v}(0) = 0$, while $\bar{u}(0)$
and $R$ are chosen self-consistently to ensure that the metric function
$u(r)$ smoothly matches onto the Schwarzschild solution $e^{2u} \to (1 - 2GM/r)$
for $r \geq R$.  In this choice of coordinates $\bar{v}(r)$ need not be
continuous at $r=R$.  The metric functions $e^{2u(r)}$ and $e^{2\bar{v}(r)}$
are depicted in Fig.~\ref{F:met}.

The final result of this
analysis is that for fixed values of $m$ and $\sigma_0$, the boson-star
solutions constitute a one-parameter family corresponding to different values
of the particle number $N$.  The boson-star mass $M$ increases monotonically
with $N$, and the boson star becomes increasingly compact.  Eventually a
critical limit of compactness $R = 2.869 GM$ is reached beyond which the boson
star collapses into a black hole.  We choose to investigate EMRIs into this
critically compact boson star.  We set $m = \sigma_0$ to the value that yields
the desired SMBS mass $M$.  For the nontopological soliton considered here,
$M \sim m_{\rm Pl}^4/m^3$, implying that a boson of mass $m \sim 10$ TeV will lead
to a SMBS of $M \sim 10^6 M_\odot$.  The SMBS mass scales differently with $m$
for other types of boson stars.

\section{Geodesics of the Boson-Star Metric} \label{S:geo}

In the extreme--mass-ratio limit $\eta \equiv \mu/M \ll 1$, where $\mu$ and
$M$ are the mass of the CO and boson star respectively, the CO will travel
along geodesics of the boson-star metric.  During the course of the EMRI,
energy and angular momentum are radiated away on timescales much longer than
an orbital period.  We can therefore make the adiabatic approximation that
the CO will migrate smoothly between geodesics characterized by decreasing energy
and angular momentum.  A thorough description of the
possible geodesics of the boson-star metric is thus essential to
understanding the evolution of the CO's trajectory.  This description can be
simplified by noting that the metric of Eq.~(\ref{E:metric}) is independent of
$t$ and $\phi$, implying the existence of timelike and azimuthal Killing
fields $\xi^{\mu}$ and $\psi^{\mu}$.  For a CO travelling along a geodesic
with four-velocity $u^\nu = (dt/d\tau, dr/d\tau, d\theta/d\tau, d\phi/d\tau)$,
the inner product of $u^\nu$ with a Killing field is
conserved \cite{wald}.  This allows a formal definition of the conserved
energy and angular momentum per unit mass,
\begin{eqnarray} \label{E:E&L}
E &\equiv& - g_{\mu\nu} \xi^{\mu} u^{\nu} = e^{2u} \frac{dt}{d\tau} \, ,
\nonumber \\
L &\equiv& g_{\mu\nu} \psi^{\mu} u^{\nu} = r^2 \frac{d\phi}{d\tau} \, .
\end{eqnarray}
Note that we can restrict ourselves to orbits with $\theta = \pi/2$ without
loss of generality because of spherical symmetry, in which case $d\theta/d\tau
= 0$.  The definitions of Eq.~(\ref{E:E&L}), coupled with the norm of the
four-velocity
\begin{equation} \label{E:4v}
g_{\mu\nu} u^{\mu} u^{\nu} = -e^{2u} \left( \frac{dt}{d\tau} \right)^2 +
e^{2\bar{v}} \left( \frac{dr}{d\tau} \right)^2 +
r^2 \left( \frac{d\phi}{d\tau} \right)^2 = -1 \, ,
\end{equation}
provide three coupled, first-order differential equations that can be solved
for the CO's orbit $\{t(\tau), r(\tau), \phi(\tau) \}$ as a function of proper
time $\tau$.

While this approach formally solves the problem of geodesic motion, further
insight can be gained by recasting Eq.~(\ref{E:4v}) to
make an analogy with one-dimensional particle motion \cite{chand},
\begin{eqnarray} \label{E:effpot}
E^2 &=& e^{2u + 2\bar{v}} \left( \frac{dr}{d\tau} \right)^2 + e^{2u}
\left( \frac{L^2}{r^2} + 1
\right) \nonumber \\
&=& e^{2u + 2\bar{v}} \left( \frac{dr}{d\tau} \right)^2 + V^2(r) \, .
\end{eqnarray}
The first term on the right-hand side of Eq.~(\ref{E:effpot}) acts like a
positive-definite kinetic energy, while the second term is the potential well
in which the one-dimensional particle motion occurs.  This analogy is not as
complete as in the case of a Schwarzschild black hole, where the metric
functions conspire to make the coefficient of the kinetic term independent of
position.  For boson stars the term $e^{2u + 2\bar{v}}$ acts like a
position-dependent mass; while this will affect the particle motion 
quantitatively, it will not preclude the existence of bound orbits in the
boson-star interior.

\begin{figure}[t!]
\scalebox{.70}{\includegraphics{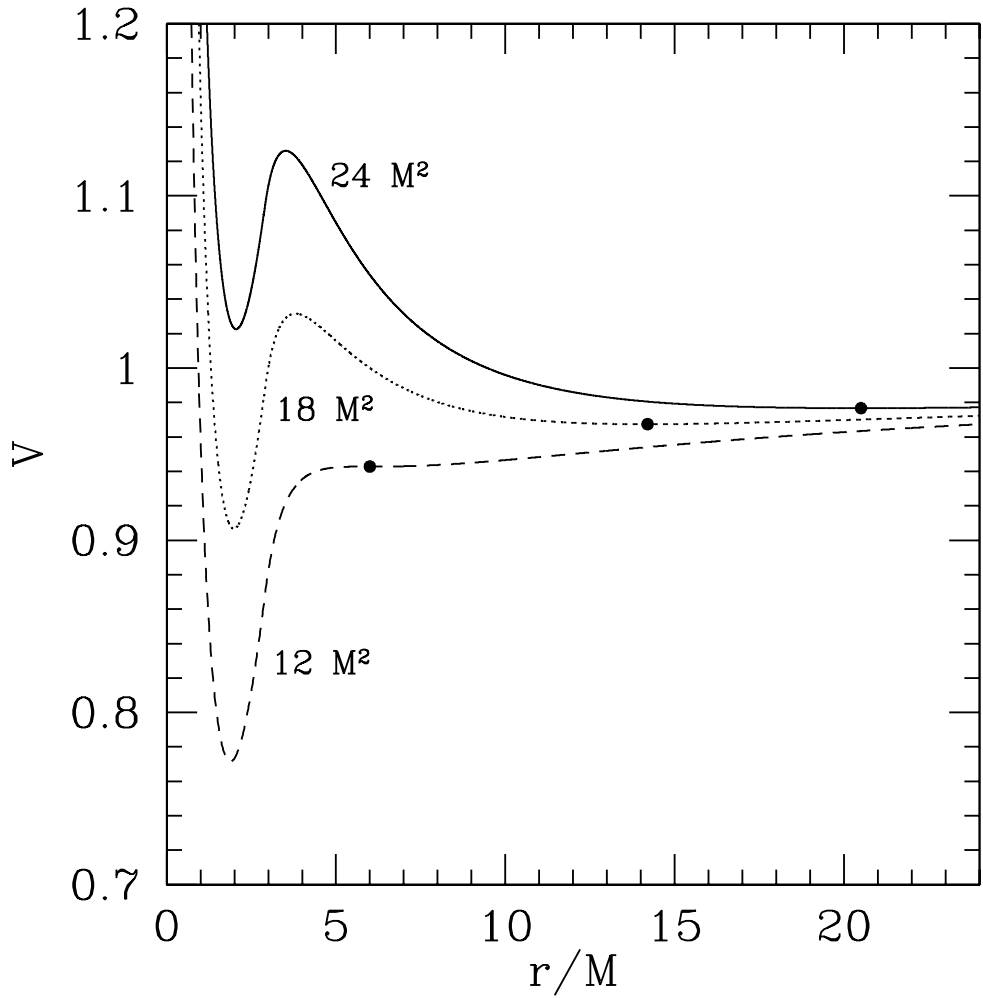}}
\caption{The effective potential $V(r)$ of the boson star for three different
values of the angular momentum.  The solid, short-dashed, and long-dashed
curves have $L^2 = 24 M^2$, $18 M^2$, and $12 M^2$ respectively.  The large
black dots on each curve show the location of the outer minimum.  For $L^2 =
12 M^2$, the lowest angular momentum for which there are two minima, the black
dot is located at the innermost stable circular orbit $r = 6M$. \label{F:pot}}
\end{figure}

Plots of the effective
potential $V(r)$ for different values of the angular momentum $L$ are given in
Fig.~\ref{F:pot}.
The effective potential per unit mass and ratio $r/M$ are
dimensionless in units where $G = c = 1$.  Outside of the boson-star surface
at $R = 2.869 M$, $V(r)$ is identical to that of Schwarzschild black holes.
For $L^2 \geq 12 M^2$, a local minimum exists at $r_1 \geq 6 M$.  Test
particles with $E = V(r_1)$ can follow stable circular orbits at $r_1$, while
those with slightly larger energies experience radial oscillations about this
minimum corresponding to eccentric orbits.  Particles with $E > 1$ are
unbound, and will either escape to infinity or, if they have energies greater
than the local maximum and $dr/dt < 0$, penetrate to the boson-star interior.
It is here that the story changes dramatically from that of the black hole,
where $V(r)$ plunges to negative infinity at the singularity.  Particles that
enter the black hole's event horizon are directed inexorably to the center,
never to return.  By contrast, the potential well of the boson star has finite
depth, implying that for $L > 0$ the effective potential is repulsive at short
distances.  This creates a second minimum in the effective potential at $r_2$
at which stable circular orbits and associated eccentric orbits can occur.
Such orbits exist even for $E > 1$, though it is unclear in what astrophysical
context a CO might find itself on such a geodesic.  For a typical EMRI, a CO
will be scattered onto an eccentric orbit and slowly circularize and wander
inward as energy and angular momentum are lost to gravitational radiation.
Eventually, either before or after circularization depending on the initial
eccentricity and semi-major axis, the CO will reach the ISO and plunge into the
boson-star interior.  At this point the trajectory and corresponding
gravitational waveform will diverge radically from the case of an EMRI into a
black hole.  Provided there is no non-gravitational interaction between the
CO and the scalar field, the inspiral will continue on a highly eccentric
geodesic about the inner minimum of the effective potential.  Energy and
angular momentum will evolve continuously through the plunge, while both the
apocenter and pericenter will change on orbital timescales as the center of
radial oscillations changes from the outer minimum $r_1$ to the inner minimum
$r_2$.  High eccentricity at the ISO leads to even more
complicated inspirals as the plunge takes place in a two-step process.
The CO crosses over the local maximum of the effective
potential several orbits before settling into the inner local minimum.  The
pericenter penetrates deep into the boson-star interior in the first stage
of this two-step plunge, while in the second stage the apocenter shifts
from outside to inside the local maximum.  This inspiral would undoubtably
produce a very distinctive waveform, but only a quantitative analysis can
determine whether such scenarios are possible. To make such a quantitative
analysis we must first develop a suitable approximation for the
calculation of the gravitational-wave emission.

\section{Gravitational Radiation} \label{S:GR}

Gravitational-wave emission by arbitrary sources is a challenging
unsolved problem.  In any asymptotically flat spacetime one can define the
GW field, 
\begin{equation} \label{E:TRdef}
\bar{h}^{\alpha\beta} \equiv -(-g)^{1/2} g^{\alpha\beta} + \eta^{\alpha\beta}
\, ,
\end{equation}
where $g_{\alpha\beta}$ is the metric, $g$ is its determinant, and
$\eta^{\alpha\beta}$ is the flat-space Lorentz metric \cite{thorne1980}.  Far
from the source, this quantity reduces to the trace-reversed metric
perturbation $\bar{h}^{\alpha\beta} = h^{\alpha\beta} - 1/2 \eta^{\alpha\beta}
h$.  The perturbation $\bar{h}^{\alpha\beta}$ satisfies the de Donder gauge
condition
\begin{equation} \label{E:dDg}
\bar{h}^{\alpha\beta},_{\beta} = 0 \, ,
\end{equation}
and the exact Einstein field equations
\begin{equation} \label{E:EFeq}
\square \bar{h}^{\alpha\beta} = -16 \pi \tau^{\alpha\beta} \, .
\end{equation}
We see that the GW field is sourced by an ``effective stress-energy tensor''
$\tau^{\alpha\beta}$ of the form \cite{thorne1980}
\begin{equation} \label{E:effstress}
\tau^{\alpha\beta} = (-g)(T^{\alpha\beta} + t^{\alpha\beta}_{\rm LL}) +
(16\pi)^{-1} [ \bar{h}^{\alpha\mu},_\nu \bar{h}^{\beta\nu},_\mu -
\bar{h}^{\alpha\beta},_{\mu\nu} \bar{h}^{\mu\nu} ] \, ,
\end{equation}
where $T^{\alpha\beta}$ is the true stress-energy tensor and
$t^{\alpha\beta}_{\rm LL}$, the Landau-Lifschitz pseudotensor given by
Eq.~(20.22) of \cite{MTW}, is a highly nonlinear function of the full metric.
We thus see that for highly relativistic, strong-field sources the GW field
$\bar{h}^{\alpha\beta}$ will itself provide a non-negligible contribution
to its source $\tau^{\alpha\beta}$.  In general Eq.~(\ref{E:EFeq}) will lead
to a system of tightly coupled, nonlinear differential equations that can only
be integrated for brief periods numerically before becoming unstable.
To make any analytical progress we must rely on a series of
simplifying approximations of varying degrees of validity.

The first approximation we make is that in the extreme--mass-ratio limit
$\eta \equiv \mu/M \ll 1$, the boson star is at rest at the origin of our
coordinate system and its metric is static.  The inspiraling CO is thus the
sole time-varying source that contributes to GW emission.  In reality the CO
should raise tides on the surface of the boson star, which could then generate
GWs themselves as well as back-react on the CO, altering its trajectory.
In an inspiral into a black hole, the energy loss due to the tidal interaction is
at greatest only a few percent of the total flux, although this can still lead to
several hundred cycles of phase difference in the gravitational waveform
\cite{finn&thorne}. In the boson-star case the nature of the tidal interaction
will be different and model-dependent, and provides another way to identify
boson-star inspirals. However, the tidal interaction should still be a lesser
effect than the orbital dynamics.
The back-reaction on the orbit is suppressed by an additional factor of $\eta$,
implying that for EMRIs with $\eta \simeq 10^{-6}$ it will introduce a very
subdominant source of error into our calculations.  The second approximation
we make is to ignore the curvature induced by the boson star as it affects the
propagation of GWs to infinity.  GWs propagating
outwards from $r \lesssim (\lambda^2 M)^{1/3}$, where $\lambda$ is the GW
wavelength, will be distorted and backscattered by the background curvature
of the boson star \cite{thorne1980}.  The importance of this effect will depend
on the inner boundary conditions, which should differ significantly from those
of a black hole with an event horizon.  We ignore this complication entirely
by assuming that backscattering does not occur.  
A third approximation that we make is to compute waveforms using
weak-field formulae. We wish to apply these to orbits that come very close
to the boson star, a regime in which there is no natural flat-space
coordinate system in which to evaluate the weak-field expressions. Our
approach is to identify the Schwarzschild coordinates ($t$, $r$, $\theta$,
$\phi$) of the inspiral trajectory with true flat-space spherical polar
coordinates. This identification is exact for orbits out in the
weak-field. There is no {\it a priori} reason why
this identification is better than any other, e.g., identifying isotropic
coordinates with the flat-space spherical polar coordinates. In practice,
using different coordinate identifications gives different answers, but
the differences are only a few percent. Taken together, these three
approximations are equivalent to assuming the GWs from the EMRI will be
the same as those that would be produced were the CO a "particle on a
string", artificially constrained to move on an orbit $\{t(\tau), r(\tau),
\phi(\tau)\}$ in flat space. In the case of black-hole inspirals,
waveforms computed using this "hybrid" approach
\cite{GHK,TSNcomp,waveform} have been found to compare
quite well to waveforms computed using more accurate perturbative
techniques.

In the flat-space, weak-field limit implied by the above approximations, the
transverse traceless (TT) part of the GW field can be expanded in symmetric
trace-free (STF) tensors \cite{thorne1980},
\begin{equation} \label{E:hMM}
h_{jk}^{\rm TT} = \left[ \sum_{l=2}^{\infty} \left( \frac{4}{l!} r^{-1} \right)
\, ^{(l)}\mathcal{I}_{jkA_{l-2}}(t-r) N_{A_{l-2}} + \sum_{l=2}^{\infty}
\left( \frac{8l}{(l+1)!} \right) r^{-1} \epsilon_{pq \left( j \right.} \,
^{(l)}\mathcal{S}_{\left. k \right) pA_{l-2}}(t-r) n_q N_{A_{l-2}}
\right]^{\rm TT} \, ,
\end{equation}
where $A_l \equiv a_1 ... a_l$, $\mathcal{I}_{A_l}$, and $\mathcal{S}_{A_l}$
are the rank-$l$ mass and current multipole-moment tensors, $n_a$ is a unit
radial vector, and $N_{A_l} \equiv n_{a_1} ... n_{a_l}$.  Parentheses on the
indices denote taking the symmetric part, and a superscript $(l)$ in front of
a tensor indicates taking the $l$-th time derivative.  The energy and angular
momentum loss to gravitational radiation can also be expanded in
multipole-moment tensors,
\begin{eqnarray}
\frac{d(\mu E)}{dt} &=& \sum_{l=2}^{\infty} \frac{(l+1)(l+2)}{(l-1)l}
\frac{1}{l!(2l+1)!!} \langle ^{(l+1)}\mathcal{I}_{A_l} \,
^{(l+1)}\mathcal{I}_{A_l} \rangle + \sum_{l=2}^{\infty} \frac{4l(l+2)}{(l-1)}
\frac{1}{(l+1)!(2l+1)!!} \langle ^{(l+1)}\mathcal{S}_{A_l} \,
^{(l+1)}\mathcal{S}_{A_l} \rangle \, , \label{E:EMM} \\
\frac{d(\mu L_j)}{dt} &=& \sum_{l=2}^{\infty}
\frac{(l+1)(l+2)}{(l-1)l!(2l+1)!!}
\langle \epsilon_{jpq} \, ^{(l)}\mathcal{I}_{pA_{l-1}} \,
^{(l+1)}\mathcal{I}_{qA_{l-1}} \rangle \nonumber \\
&& \quad + \sum_{l=2}^{\infty}
\frac{4l^2(l+2)}{(l-1)(l+1)!(2l+1)!!} \langle \epsilon_{jpq} \,
^{(l)}\mathcal{S}_{pA_{l-1}} \, ^{(l+1)}\mathcal{S}_{qA_{l-1}} \rangle \, .
\label{E:LMM}
\end{eqnarray}
Here $\langle \quad \rangle$ denotes averaging over an entire orbital period.
A $\mu$ appears on the left-hand side of the equation in accordance with our
definition of $E$ and $L$ as the energy and angular momentum per unit mass.
This multipole-moment expansion is valid in principle for sources moving at
arbitrarily relativistic velocities.  We now further approximate that the
sources are Newtonian; the maximum velocity is only mildly relativistic, and
the internal stresses are small compared to the energy density
\cite{thorne1980}.  Although this approximation is violated during the EMRI in
the vicinity of the plunge, key features of the relativistic motion are
preserved by constraining the CO to travel along geodesics that are {\it exact}
even in the relativistic regime.  Empirically, expansions
in the post-Newtonian parameter $(L/\lambda)^2 \sim v^2$ are found to give
physical results (no outspiral) even in regions where they do not converge
\cite{BC1}.  Assuming Newtonian sources, the mass and current multipole tensors
are given by \cite{thorne1980}
\begin{eqnarray}
\mathcal{I}_{A_l} &=& \left[ \int \rho X_{A_l} \, d^3 \vec{x} \right]^{\rm STF}
\, , \label{E:Inewt} \\
\mathcal{S}_{A_l} &=& \left[ \int ( \epsilon_{a_l pq} x_p \rho v_q )
X_{A_{l-1}} \, d^3 \vec{x} \right]^{\rm STF} \, , \label{E:Snewt}
\end{eqnarray}
where $X_{A_l} \equiv x_{a_1} ... x_{a_l}$, and $\rho$ is the energy density.
In keeping with our assumption that the static boson-star metric does not
contribute to the production of GWs, the energy density is simply that of the
CO,
\begin{equation} \label{E:COed}
\rho(\vec{x}) = \mu \delta^3 (\vec{x} - \vec{x}_{\rm CO}(t)) \, ,
\end{equation}
where $\vec{x}_{\rm CO}(t)$ is the flat-space trajectory of the CO.  The mass
and current multipole moments $\mathcal{I_{A_l}}$ and $\mathcal{S_{A_l}}$
contribute to the GW field $h_{jk}^{\rm TT}$ at order $(M/r)(L/\lambda)^l$
and $(M/r)(L/\lambda)^{l+1}$ respectively, while the assumption of Newtonian
sources induces errors of order $(L/\lambda)^2$ to each term
\cite{thorne1980}.  It is therefore inconsistent to include terms higher than
the mass quadrupole, current quadrupole, and mass octupole under this
assumption.  We consider only the mass quadrupole term, relegating an analytic
calculation of the error associated with this approximation to the Appendix.
In this case,
Eqs.~(\ref{E:hMM}), (\ref{E:EMM}), and (\ref{E:LMM}) reduce to the familiar
form of the ``quadrupole approximation''
\begin{eqnarray}
h_{jk}^{\rm TT} &=& \frac{2}{r} \frac{d^2 \mathcal{I}_{jk}}{dt^2} \, ,
\label{E:h2} \\
\frac{d(\mu E)}{dt} &=& \frac{1}{5} \left< \frac{d^3 \mathcal{I}_{jk}}{dt^3}
\frac{d^3 \mathcal{I}_{jk}}{dt^3} \right> \, , \label{E:E2} \\
\frac{d(\mu L_j)}{dt} &=& \frac{2}{5} \left< \epsilon_{jpq}
\frac{d^2 \mathcal{I}_{pk}}{dt^2}
\frac{d^3 \mathcal{I}_{qk}}{dt^3} \right> \, . \label{E:L2}
\end{eqnarray}
How can the humble quadrupole approximation be justified for the eccentric,
highly relativistic final stages of an EMRI into a boson star?  While the
energy and angular momentum fluxes are indeed only approximate, the
derivatives relating these fluxes to changes in the orbital parameters are
{\it exact}, as are the equations of
motion used in performing the orbital averages of Eqs.~(\ref{E:E2}) and
(\ref{E:L2}).  Unlike direct post-Newtonian expansions for the time evolution 
of orbital elements, this ``hybrid approximation'' incorporates the exact
orbital dynamics of geodesic motion \cite{GHK,PS2}. This approach is
self-consistent in the sense that the energy and angular momentum carried away by
gravitational waves (within our quadrupole approximation) is
equal to the loss of angular momentum and energy of the
orbit.  This approach can reproduce features missing from direct
post-Newtonian expansions (which conserve energy and angular
momentum only to ${\cal O}(v^2)$) such as the {\it increase}
in orbital eccentricity just prior to plunge.
Moreover, using exact geodesics in the flat-space
quadrupole formula ensures that the fundamental frequencies of the orbit
are reproduced in the gravitational waveforms. Although the distribution
of power between harmonics is not correct, these approximate waveforms do
encode the same orbital dynamics as the true waveform and therefore the
qualitative features of inspiral waveforms should be well represented.
At substantial computational
expense, gravitational perturbation theory can calculate waveforms at infinity
using the Teukolsky-Sasaki-Nakamura (TSN) formalism \cite{teuk,SN}.  Such
calculations for EMRIs in the Schwarzschild spacetime show agreement with the
hybrid approach to within 5 to $45\%$ for the time derivatives of the orbital
semi-latus rectum and eccentricity even for moderate eccentricity $(e \sim 0.4)$
at $r \simeq 7M$ \cite{GHK,TSNcomp}.  We hope that such accuracy will be
retained in the case of boson stars, sparing us for now the additional
computational expense of a numerical TSN approach.  The
details of adapting this hybrid approximation to boson-star EMRIs is the
subject of the next Section.

\section{Orbital Evolution} \label{S:OE}

In a spherically symmetric spacetime, conservation of angular momentum implies
that geodesic motion will be confined to an orbital plane \cite{SSD}.  Without
loss of generality, this plane can be chosen to be the equatorial plane
$\theta = \pi/2$.  In this case, the position of the CO is fully specified by
a radius $r$ and a true longitude $\phi$.  A geodesic will be
characterized by orbital elements like the semi-latus rectum $p$ and
eccentricity $e$, in terms of which the radius is given by
\begin{equation} \label{E:TA}
r = \frac{p}{1 + e \cos \psi} \, .
\end{equation}
The true anomaly $\psi$ can be specified independently of $\phi$, since in
general the pericenter will precess.  In the hybrid approximation, EMRI orbits
and waveforms are calculated in a two-step process that hinges on the
assumption of adiabaticity; i.e. orbital elements like $p$ and $e$ vary on
timescales much longer than an orbital period over which $\psi$ and $\phi$ change
\cite{GHK,PS2}.  This allows the differential equations governing the evolution
of $p$ and $e$ to be integrated with a much longer timestep than is required for
evolving $\psi$ and $\phi$.  We will first describe how the two stages of
integration are accomplished with this assumption, and then
we will discuss the corrections necessary to patch together a physically
reasonable orbit through those regions where the assumption is violated.

In the first stage of our calculation, the EMRI's trajectory through
a phase space of $p$ and $e$ is determined by relating these quantities to $E$
and $L$, and then using this relation and Eqs.~(\ref{E:E2}) and (\ref{E:L2})
to obtain differential equations for $p$ and $e$.  Particular values of
$p$ and $e$ uniquely determine the pericenter $r_p = p/(1+e)$ and apocenter
$r_a = p/(1-e)$ from Eq.~(\ref{E:TA}).
At both pericenter and apocenter, $dr/d\tau = 0$ implying $V(r_p) = V(r_a) = E$
by Eq.~(\ref{E:effpot}).  This equation can be solved to obtain the energy
$E$ and $L$ as functions of $p$ and $e$,
\begin{eqnarray} \label{E:EL(pe)}
L(p,e) &=& \left[ \left( e^{2u(r_p)} - e^{2u(r_a)} \right)
\left( \frac{e^{2u(r_a)}}{r_{a}^2} - \frac{e^{2u(r_p)}}{r_{p}^2} \right)^{-1}
\right]^{1/2} \, , \nonumber \\
E(p,e) &=&  \left[ e^{2u(r_a)} \left( \frac{L^2}{r_{a}^2} + 1 \right)
\right]^{1/2} \, .
\end{eqnarray}
First-order differential equations
for the time evolution of $p$ and $e$ can then be derived from the {\it exact}
derivatives of $E(p,e)$ and $L(p,e)$ along with the quadrupole appoximations
of Eqs.~(\ref{E:E2}) and (\ref{E:L2}),
\begin{eqnarray} \label{E:dpde}
\frac{dp}{dt} &=& \frac{\frac{dL}{dt} -
\frac{\partial L/\partial e}{\partial E/\partial e} \frac{dE}{dt}}{\frac{
\partial L}{\partial p} - \frac{\partial L/\partial e}{\partial E/\partial e}
\frac{\partial E}{\partial p}} \, , \nonumber \\
\frac{de}{dt} &=& \frac{\frac{dL}{dt} -
\frac{\partial L/\partial p}{\partial E/\partial p} \frac{dE}{dt}}{\frac{
\partial L}{\partial e} - \frac{\partial L/\partial p}{\partial E/\partial p}
\frac{\partial E}{\partial e}} \, .
\end{eqnarray}
It is the combination of these exact derivatives with the quadrupole
approximation for the fluxes that accounts for the superior performance of this
``hybrid'' approximation over the consistent (in that we include {\it all} terms
up to a given order in $v^2$) post-Newtonian approach.  The
quadrupole fluxes themselves can be determined by inserting Eq.~(\ref{E:COed})
for the CO's energy density into Eq.~(\ref{E:Inewt}) for the Newtonian
quadrupole moments.  Using an overdot to denote a derivative with
respect to coordinate time $t$, we find that the time derivatives of the
quadrupole moment appearing in Eqs.~(\ref{E:E2}) and (\ref{E:L2}) take the form
\begin{eqnarray} \label{E:Ider}
\mathcal{\dddot{I}}_{jk} \mathcal{\dddot{I}}_{jk} &=& 2 \eta^2 [ (1/3) (\dddot{r}
+ 3\ddot{r}\dot{r})^2 + (6\ddot{r}\dot{\phi}r + 6\dot{r}\ddot{\phi}r
+ 6\dot{r}^2 \dot{\phi} - 4\dot{\phi}^3 r^2 + \dddot{\phi}r^2)^2 + (\dddot{r}r
+ 3\ddot{r}\dot{r} - 12 \dot{r}\dot{\phi}^2 r - 6 \ddot{\phi}\dot{\phi}r^2)^2
] \, , \nonumber \\
\epsilon_{jpq} \mathcal{\ddot{I}}_{pk} \mathcal{\dddot{I}}_{qk} &=& 2 \eta^2
[ (\ddot{r}r + \dot{r}^2 -2r^2 \dot{\phi}^2)(6\ddot{r}\dot{\phi}r
+ 6\dot{r}\ddot{\phi}r + 6\dot{r}^2 \dot{\phi} - 4\dot{\phi}^3 r^2
+ \dddot{\phi}r^2) \nonumber \\
&& \quad - ( 4\dot{r}r\dot{\phi} + r^2 \ddot{\phi} ) (\dddot{r}r
+ 3\ddot{r}\dot{r} - 12 \dot{r}\dot{\phi}^2 r - 6 \ddot{\phi}\dot{\phi}r^2)
] \, .
\end{eqnarray}
The derivatives appearing in Eq.~(\ref{E:Ider}) can be evaluated analytically
by solving Eqs.~(\ref{E:E&L}) and (\ref{E:4v}) for the equations of motion
\begin{eqnarray}
\frac{dt}{d\tau} &=& Ee^{-2u} \, , \label{E:dt} \\
\frac{d\phi}{d\tau} &=& \frac{L}{r^2} \, , \label{E:dphi} \\
\left( \frac{dr}{d\tau} \right) &=& e^{-\bar{2v}} \left[ E^2 e^{-2u} - \left(
\frac{L^2}{r^2} + 1 \right) \right] \, , \label{E:dr}
\end{eqnarray}
dividing Eqs.~(\ref{E:dr}) and (\ref{E:dphi}) by Eq.~(\ref{E:dt}), and then
taking the appropriate derivatives.  We then perform the orbital average
appearing in Eqs.~(\ref{E:E2}) and (\ref{E:L2}),
\begin{equation} \label{E:ave}
\langle f(r) \rangle = \frac{ \int_{r_p}^{r_a} f(r) \, \frac{dr}{\dot{r}}}
{\int_{r_p}^{r_a} \frac{dr}{\dot{r}}} \, .
\end{equation}
The products of these manipulations are first-order differential equations for
$\dot{p}$ and $\dot{e}$ that can be integrated forward in time for arbitrary
initial $p_0$ and $e_0$.  This tabulated phase-space trajectory
$\{ p(t), e(t) \}$ will then be used in the second stage to produce the
real-space trajectory $\{ r(t), \phi(t) \}$.

The second stage of calculating the EMRI real-space trajectory is essentially
no more complicated than integrating the equations of motion Eqs.~(\ref{E:dt}),
(\ref{E:dphi}), and (\ref{E:dr}) with a timestep that is a sufficiently small
fraction of the orbital period as to obtain the desired accuracy.  The
phase-trajectory determined in the first stage can be linearly interpolated to
obtain $p(t)$ and $e(t)$, from which the energy and angular momentum follow
from the relations $E(p,e)$ and $L(p,e)$.  These are then inserted into the
equations of motion along with $r$ and $\phi$ at the beginning of the timestep.
A minor technical difficulty arises for circular orbits $(e = 0)$ and at the
turning points of eccentric orbits $(\psi = n\pi)$ where the relation
\begin{equation}
\frac{d\psi}{d\tau} = \frac{1 + e\cos \psi}{er\sin \psi} \frac{dr}{d\tau}
\end{equation}
becomes undefined.  This problem can be solved by a judicious Taylor expansion
of Eq.~(\ref{E:dr}) about the appropriate $r$ values.  Once the real-space
trajectory $\{ x(t), y(t), z(t) \}$ has been obtained, the quadrupole waveform
of Eq.~(\ref{E:h2}) can be calculated by taking the appropriate numerical
derivatives.  This portion of the calculation does not involve any special
features of the boson-star spacetime, in keeping with the ``particle on a
string'' approximation.  The GW field $h_{ij}^{\rm TT}$ can then be
decomposed into plus and cross polarizations by defining unit vectors in the
plane of the sky.  With $\hat{L}$ a unit vector parallel to the CO's angular
momentum and $\hat{n}$ a unit vector pointed from the observer to the source,
we define
\begin{eqnarray} \label{E:BV}
\hat{p} &\equiv& (\hat{n} \times \hat{L})/|\hat{n} \times \hat{L}| \, ,
\nonumber \\
\hat{q} &\equiv& \hat{n} \times \hat{p} \, .
\end{eqnarray}
Note that the sign of $\hat{q}$ differs from that of
Ref. \cite{BC1}, but in other
respects our conventions are equivalent.  Our basis vectors $\hat{p}$ and
$\hat{q}$ are constants because $\hat{L}$ does not precess in a spherically
symmetric spacetime.  The two polarization basis tensors can be defined as
\begin{eqnarray} \label{E:BT}
H_{ij}^+ &\equiv& \hat{p_i} \hat{p_j} - \hat{p_i} \hat{p_j} \, ,
\nonumber \\
H_{ij}^\times &\equiv& \hat{p_i} \hat{q_j} + \hat{q_i} \hat{p_j} \, ,
\end{eqnarray}
allowing the GW field to be expressed as two amplitudes,
\begin{equation} \label{E:hA}
h_{ij}^{\rm TT}(t) = A^+(t) H_{ij}^+ + A^\times(t) H_{ij}^\times \, .
\end{equation}
In the next Section, we will examine the trajectories $\{ x(t), y(t), z(t) \}$
and GW amplitudes $\{ A^+(t), A^\times(t) \}$ for different EMRIs.

Before presenting these results, we must explain the kludges we have chosen
where the hybrid approximation described above is no longer valid.  Our
philosophy is to adopt the simplest, least computationally intensive approach
that yields physically reasonable waveforms.  The numerous approximations
involved in the ``particle on a string'' model introduce at least a $10 \%$
error into our trajectories and waveforms in the highly eccentric, relativistic
limit \cite{GHK,TSNcomp}; to demand greater accuracy where the hybrid
approximation is also violated would be inconsistent.  Special provisions beyond
the hybrid approximation must be made whenever $p$ and $e$ are
changing on timescales shorter than the orbital period.  For extremely
eccentric orbits, gravitational radiation will tend to be emitted in bursts
near pericenter, while the CO will be found near apocenter for most of the
orbital period.  This is particularly true of ``zoom-whirl'' orbits like those
found around rapidly spinning black holes, where the CO may whirl for many
radians near pericenter before zooming back out to apocenter \cite{zoom}.  In
such cases, the orbital averaging of Eq.~(\ref{E:ave}) will lead to an
artificially smooth phase-space trajectory $\{ p(t), e(t) \}$.  For boson-star
EMRIs, this adiabatic approximation is most flagrantly violated during the
plunge itself.  Since the rapidly varying true anomaly $\psi(t)$ is only
calculated in the second stage of our two-step approach, the CO could be
anywhere along its orbit at the time the plunge is calculated to occur in the
first stage.  This poses problems in both stages of the calculation.  As
described qualitatively in \S \ref{S:geo}, a CO initially oscillating about the
outer minimum of the effective potential $V(r)$ can plunge into the boson star
with $L^2 > 12 M^2$, in which case two minima will still exist even after the
plunge.  For a time, the energy of the CO will exceed $V_{\rm max}(L)$, the
local maximum of $V(r)$, so that {\it both} of the local minima will lie
between the orbit's pericenter and apocenter.  Such a highly eccentric orbit
will radiate energy very efficiently, implying that the CO will quickly plunge
into the inner minimum.  However, if the initial conditions were such that the CO
began on a geodesic with $E$ was significantly above $V_{\rm max}$, there is no
reason {\it a priori} why it could not first fall ino the outer minimum before
eventually inspiraling into the boson-star interior.  Which minimum
the CO ends up in depends on where the CO is in its orbit when $E$ drops below
$V_{\rm max}$.  But the true anomaly $\psi(t)$ is not calculated in the first
stage of the hybrid approximation, so we have no way of accurately knowing
which minimum to choose.  The semi-latus rectum $p$ and eccentricity $e$
change discontinuously when the CO falls into one of the two minima, a sure sign
that the adiabatic condition no longer holds.  We assume without rigorous proof
that the CO {\it always} ends up in the inner minimum in such a situation.  This
assumption comes back to haunt us in the
second stage, since the CO may well be near apocenter at the time determined for
the plunge in the first stage.  In this case, no possible instantaneous change
in $\psi$ would allow $r$ to remain continuous.  Instead, we fix $p$ and $e$ to
their values at the time of the plunge, and allow the CO to travel along a
geodesic until it reaches the local maximum at $r_{\rm max}$.  This then
becomes the apocenter of the CO's radial
motion about the inner minimum, and we resume linearly interpolating $p$ and $e$
from the phase-space trajectory tabulated in the first stage.  Having
exhaustively described our technique, we will now examine the fruits of our
labor.

\section{Results} \label{S:res}

\begin{figure}[t!]
\scalebox{.50}{\includegraphics{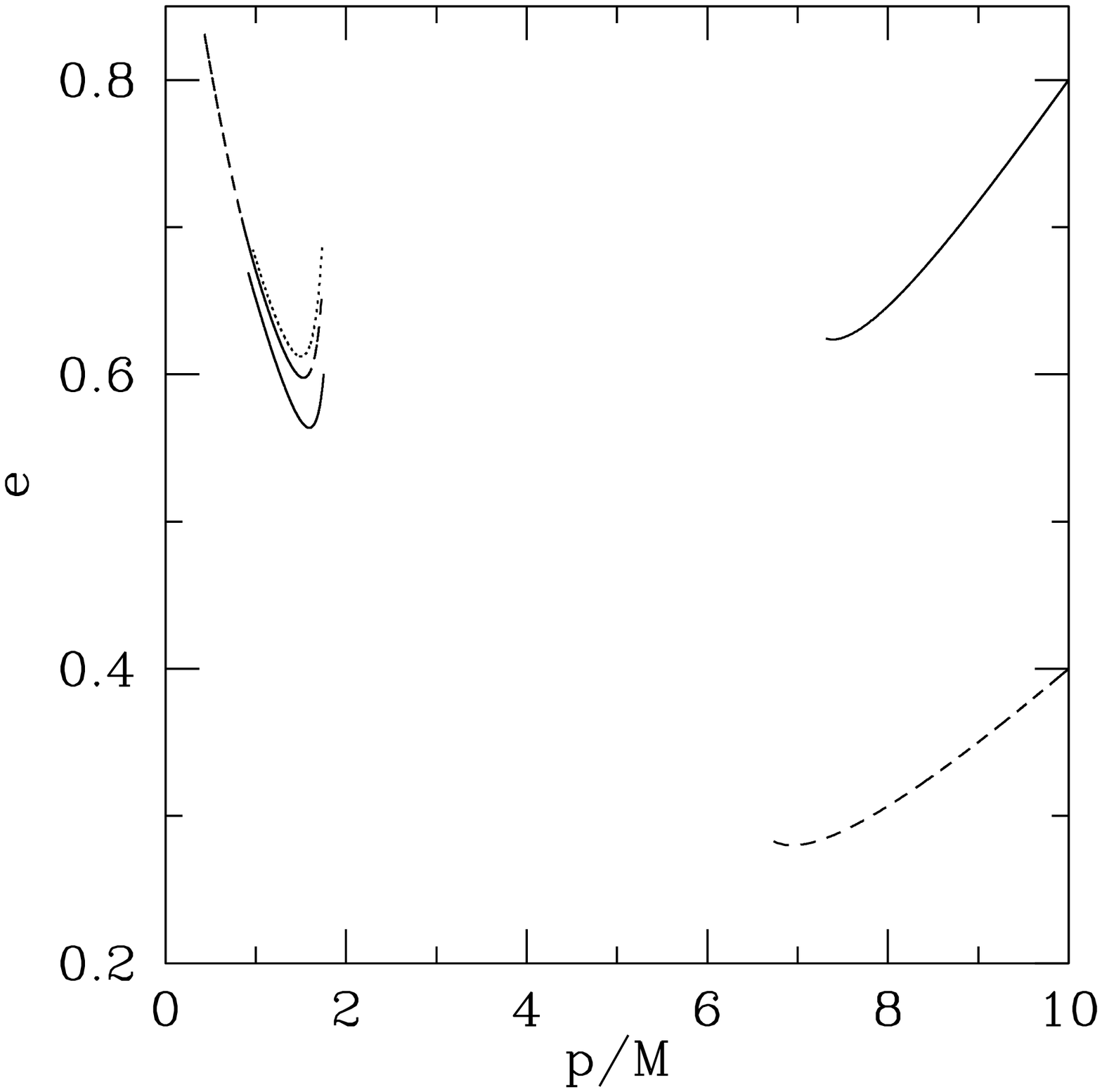}}
\scalebox{.50}{\includegraphics{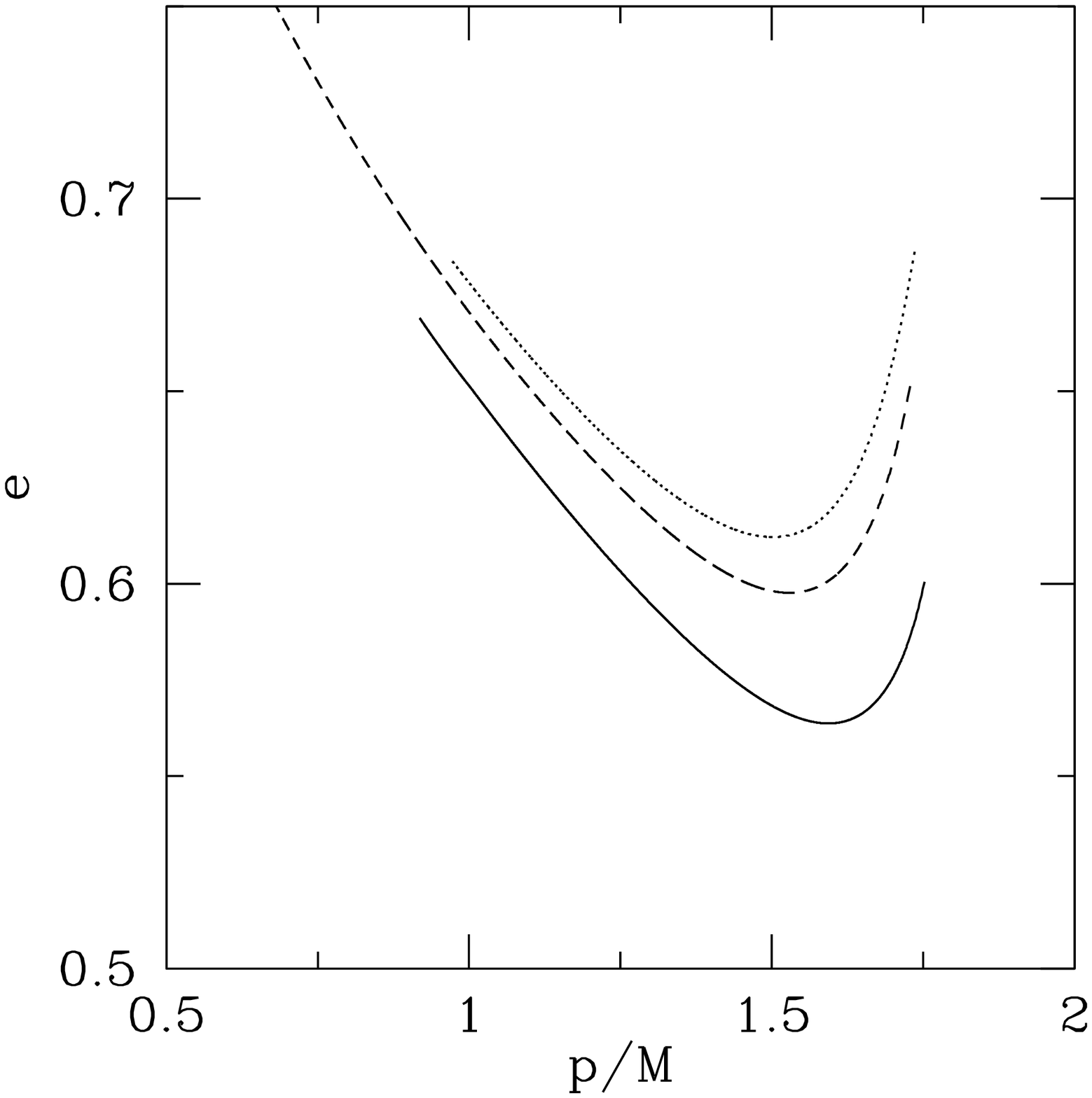}}
\caption{The phase-space trajectories $\{ p(t), e(t) \}$ for three different
EMRIs into a supermassive boson star.  The solid, dashed, and dotted
curves begin with eccentricities $e_0 = 0.8$, 0.4, and 0.0 respectively.  The
top panel shows the complete EMRI from $p_0/M = 10.0$, while the lower panel
is a close-up of the post-plunge phase of the EMRI that appears in the
upper-left corner of the top panel.  In both panels time increases from right to
left as $p/M$ decreases monotonically.
\label{F:pe}}
\end{figure}

\begin{figure}[t!]
\scalebox{.50}{\includegraphics{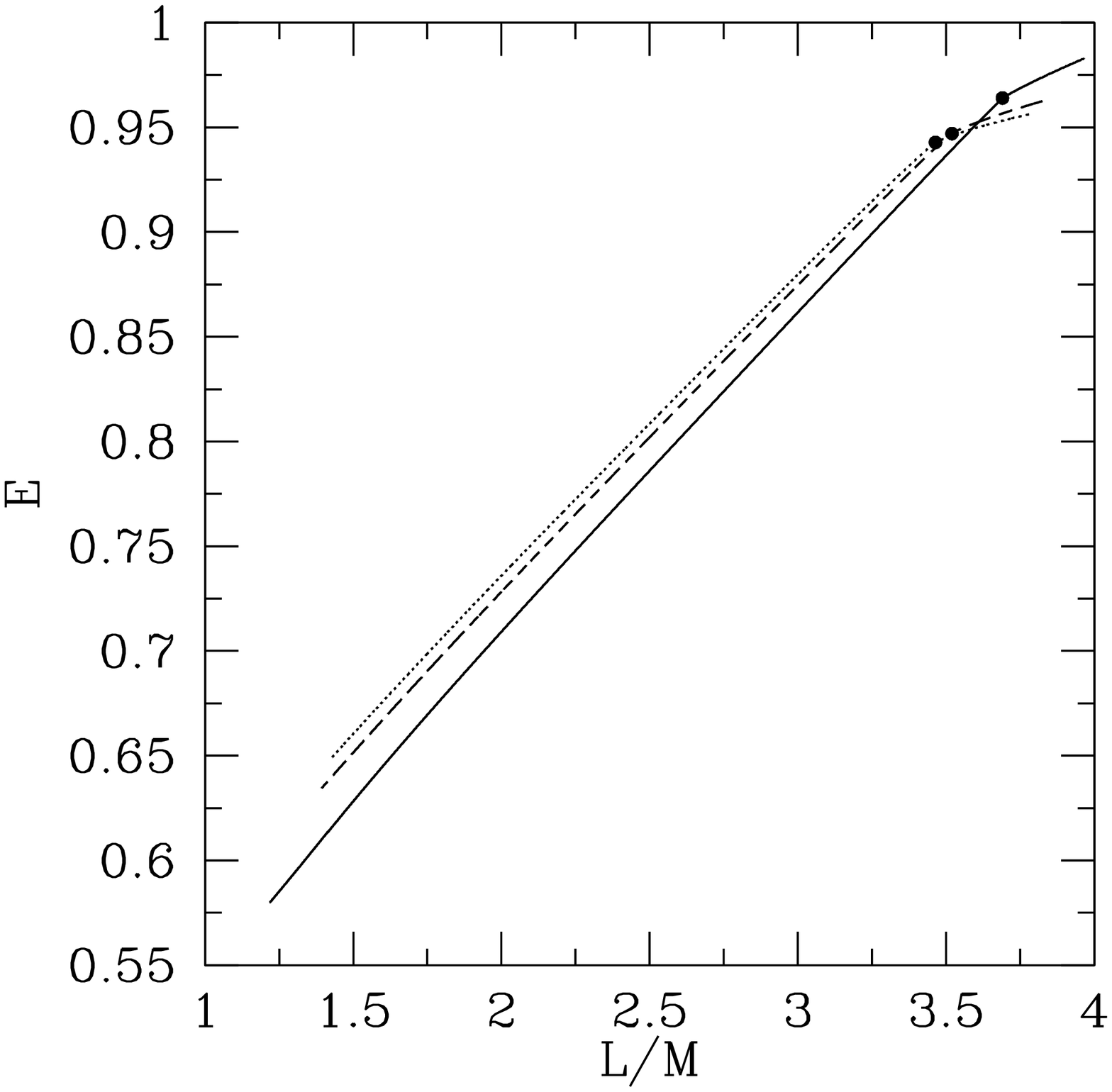}}
\caption{The phase-space trajectories $\{ L(t), E(t) \}$ for the three EMRIs
depicted in Fig.~\ref{F:pe}.  As before, the solid, dashed, and dotted
curves begin with eccentricities $e_0 = 0.8$, 0.4, and 0.0 respectively.  The
large black dots correspond to the points at which the COs plunge into the boson
star.  Note that the curves cross each other, implying that two different
geodesics can be characterized by the {\it same} energy and angular momentum.
This feature, unique to the boson-star case, follows from the existence of
bound geodesics both interior and exterior to the local maximum of the effective
potential for certain energy and angular momenta.  No such crossing appear in
Fig.~\ref{F:pe} as $p$ and $e$ do uniquely specify a given geodesic.
\label{F:EL}}
\end{figure}

First we consider the phase-space trajectories $\{ p(t), e(t) \}$ generated in
the first part of our calculation.  We choose for our fiducial model a
$3 M_\odot$ CO spiraling into a $3 \times 10^6 M_\odot$ boson star.  Three
different trajectories are depicted in Fig.~\ref{F:pe}, each with $p_0 = 10 M$
but with differing initial eccentricities $(e_0 = 0.0, 0.4, 0.8)$.
Several features of the phase-space evolution are particularly striking.
Outside the boson-star surface at $R = 2.869 GM$, the spacetime is identical to
that of a Schwarzschild black hole and the behavior of the EMRI is well known.
Orbits far from the boson star rapidly circularize, but those that retain an
appreciable eccentricity exhibit an increase in the eccentricity in the last
few orbits before the ISO.  This behavior
reflects the increasing shallowness of the outer minima as the ISO is
approached, and is not captured by post-Newtonian expansions that ignore
geodesic motion \cite{GHK,zoom}.  The second important feature is the
discontinuity in each of the three trajectories as the plunge is reached.  As
described in the previous Section, the actual position of the CO remains
continuous; only the orbital elements $p$ and $e$ change at the instant of the
plunge.  This is accomplished by an appropriate instantaneous change in the
true anomaly $\psi$.  For the EMRIs with $e_0 = 0.8$ and 0.4, the orbit is
actually divided into three distinct pieces, with the middle section
belonging to geodesics where {\it both} minima lie between pericenter and
apocenter.
This portion cannot be seen in Fig.~\ref{F:pe} because it only lasts for a few
orbital periods, which is short compared to the time scales on which $p$ and
$e$ are changing.  It is interesting to note that the three curves
have reversed their order in the plunge; the more eccentric the EMRI prior to
plunge, the lower its eccentricity afterwards.  This can be understood by
realizing that orbits with higher residual eccentricity plunge into the boson
star with higher angular momentum $L$, implying that the inner minimum will be
much steeper as depicted in Fig.~\ref{F:pot}.  The pericenter and apocenter
will be closer together in this narrow potential well, leading to a
correspondingly smaller post-plunge eccentricity.
A final surprising feature of these EMRIs is the sharp
increase in eccentricity for the final stages of the EMRI deep within the
interior of the boson star.  The energy and angular momentum are monotonically
decreasing during this stage of the EMRI in keeping with GW-induced losses;
only the relations $E(p,e)$ and $L(p,e)$ are unusual.  As angular momentum is
lost, the narrow inner minimum rapidly broadens, leading to a greater separation
between pericenter and apocenter.  The separation
\begin{equation} \label{E:sep}
r_a - r_p = \frac{2pe}{1 - e^2}
\end{equation}
is a sharply increasing function of eccentricity, explaining why the broadening
inner minimum leads to rising eccentricity.  For completeness, a plot of these
same trajectories in the phase space of angular momentum and energy
$\{ L(t)/M, E(t) \}$ is given in Fig.~\ref{F:EL}.
Note that $E$ and $L$ are continuous at the plunge as is required for physical
quantities.

\begin{figure}[t!]
\scalebox{.50}{\includegraphics{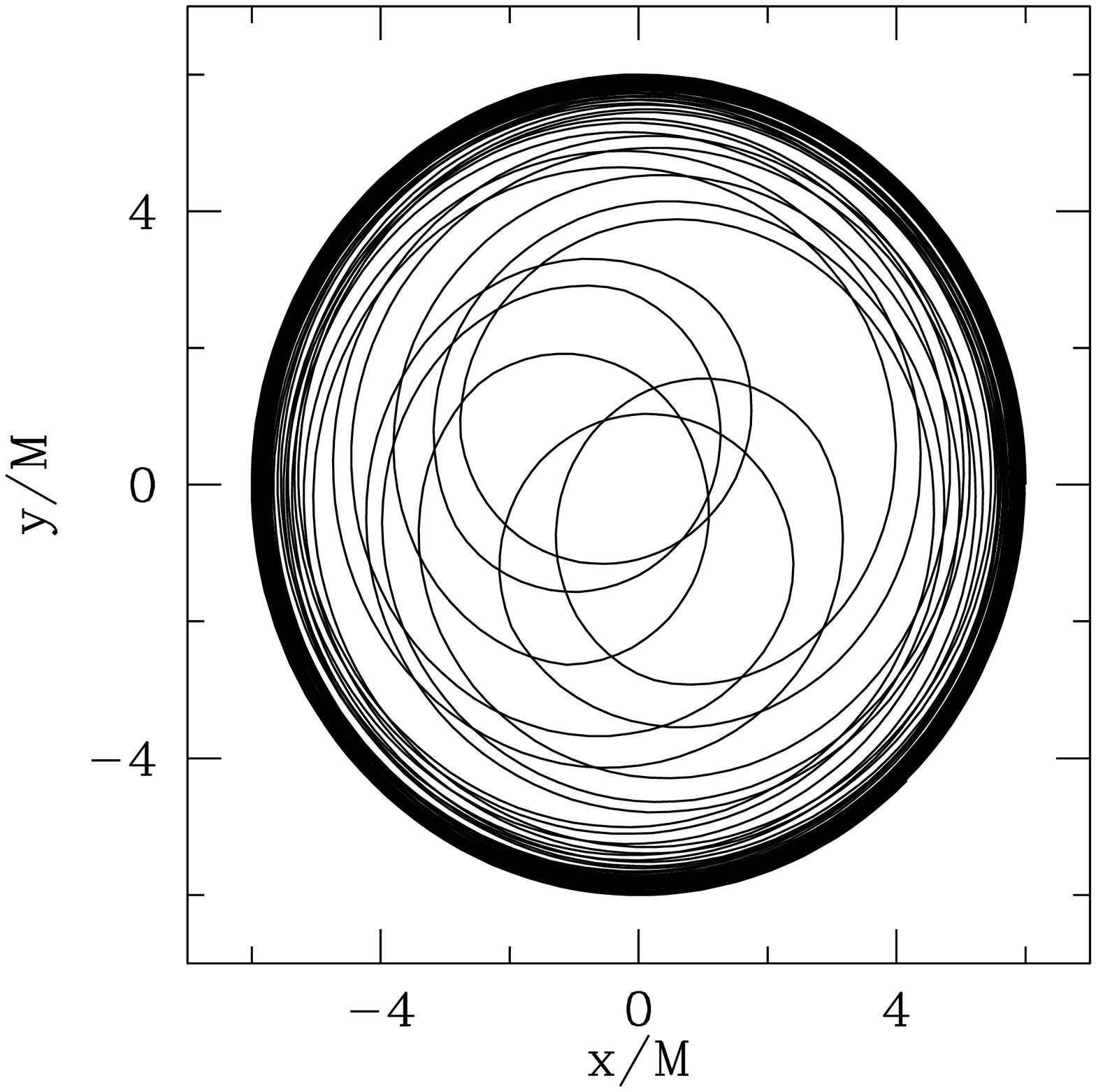}}
\caption{The real-space trajectory $\{ x(t), y(t) \}$ for an initially
circular EMRI.  For purposes of clarity, only a
period of approximately 150,000 s in the vicinity of the plunge is depicted.
\label{F:xcirc}}
\end{figure}

\begin{figure}[t!]
\scalebox{.50}{\includegraphics{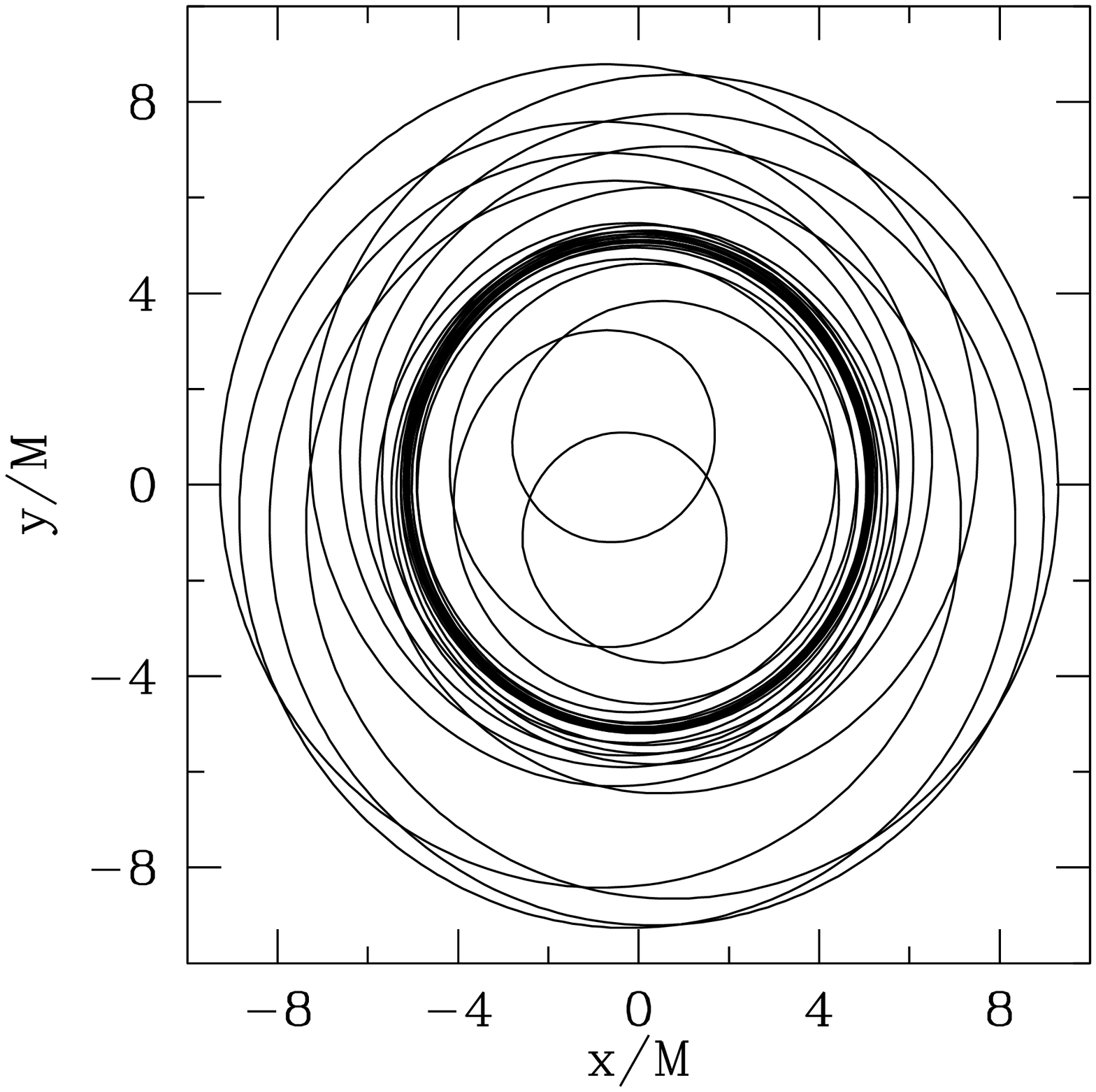}}
\scalebox{.50}{\includegraphics{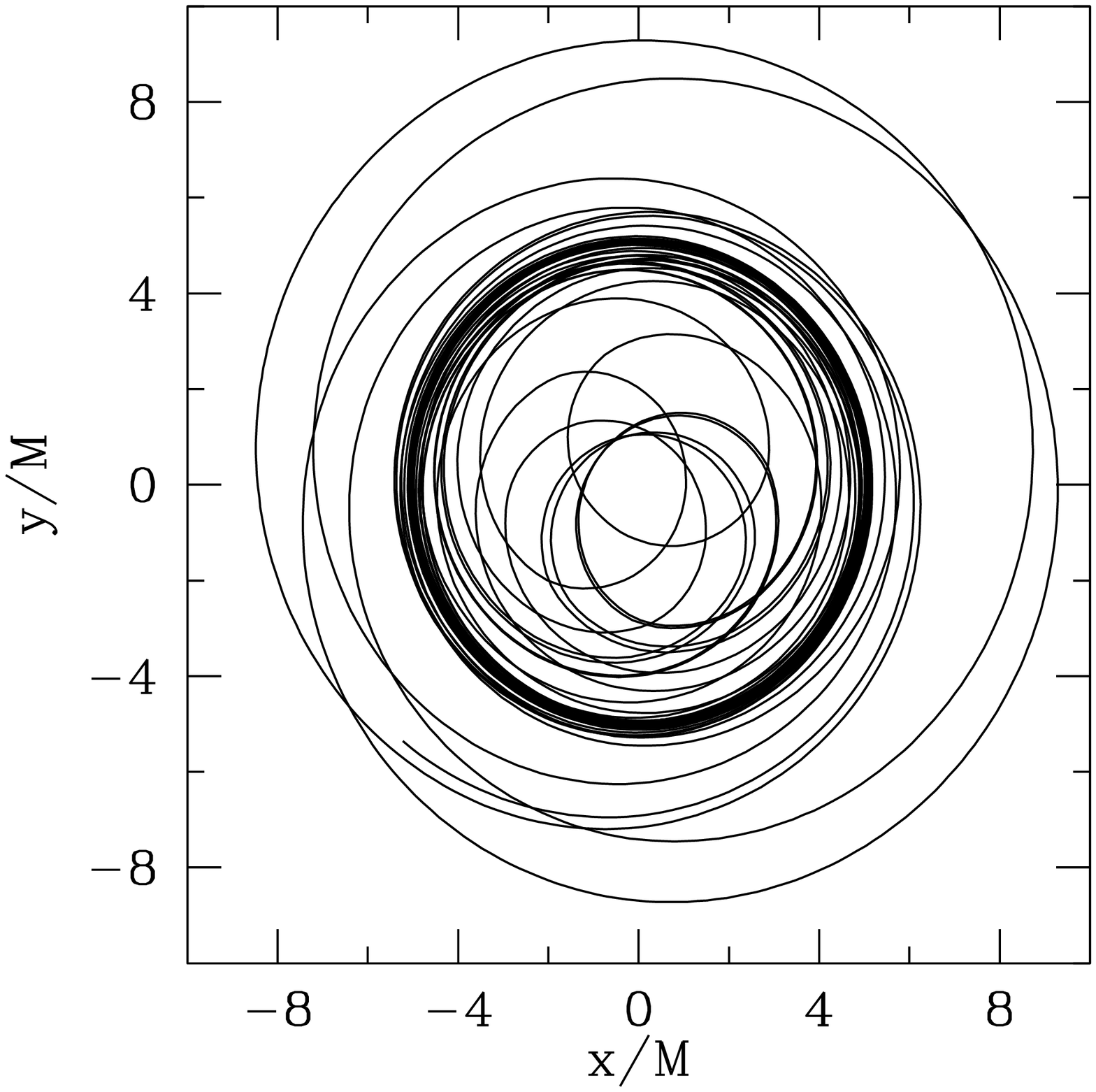}}
\caption{The real-space trajectory $\{ x(t), y(t) \}$ for the EMRI with
$e_0 = 0.4, p_0/M = 10.0$.  The upper panel shows 40,000 s including the
first stage of the two-step plunge, while the lower panel shows about 50,000 s
including the second stage.
\label{F:xecc}}
\end{figure}

We now consider the real-space trajectory $\{ x(t), y(t), z(t) \}$ itself.
Choosing the orbital and equatorial planes to coincide, $z = 0$.  Several orbits
of an initially circular EMRI on both sides of the plunge are shown in
Fig.~\ref{F:xcirc}. 
The trajectory remains continuous during the plunge, and the pericenter
precesses by an appreciable fraction of a radian on each post-plunge orbit.  Of
more interest is the EMRI of Fig.~\ref{F:xecc} with $e_0 = 0.4, p_0 = 10.0 M$.
We have omitted the EMRI with $e_0 = 0.8$ because the two initially eccentric
EMRIs appear qualitatively similar.
Both the $e_0 = 0.4$ and $e_0 = 0.8$ EMRIs experience a two-step plunge as
described qualitatively in \S \ref{S:geo}.  In the first stage, the pericenter
migrates deep into the interior of the boson star.  Only several orbits later
does the apocenter finally move inward of the local maximum.  While two-step
plunge may not be immediately apparent from the real-space trajectory, it
leaves a very distinctive signature in the gravitational waveform.

\begin{figure}[t!]
\scalebox{.50}{\includegraphics{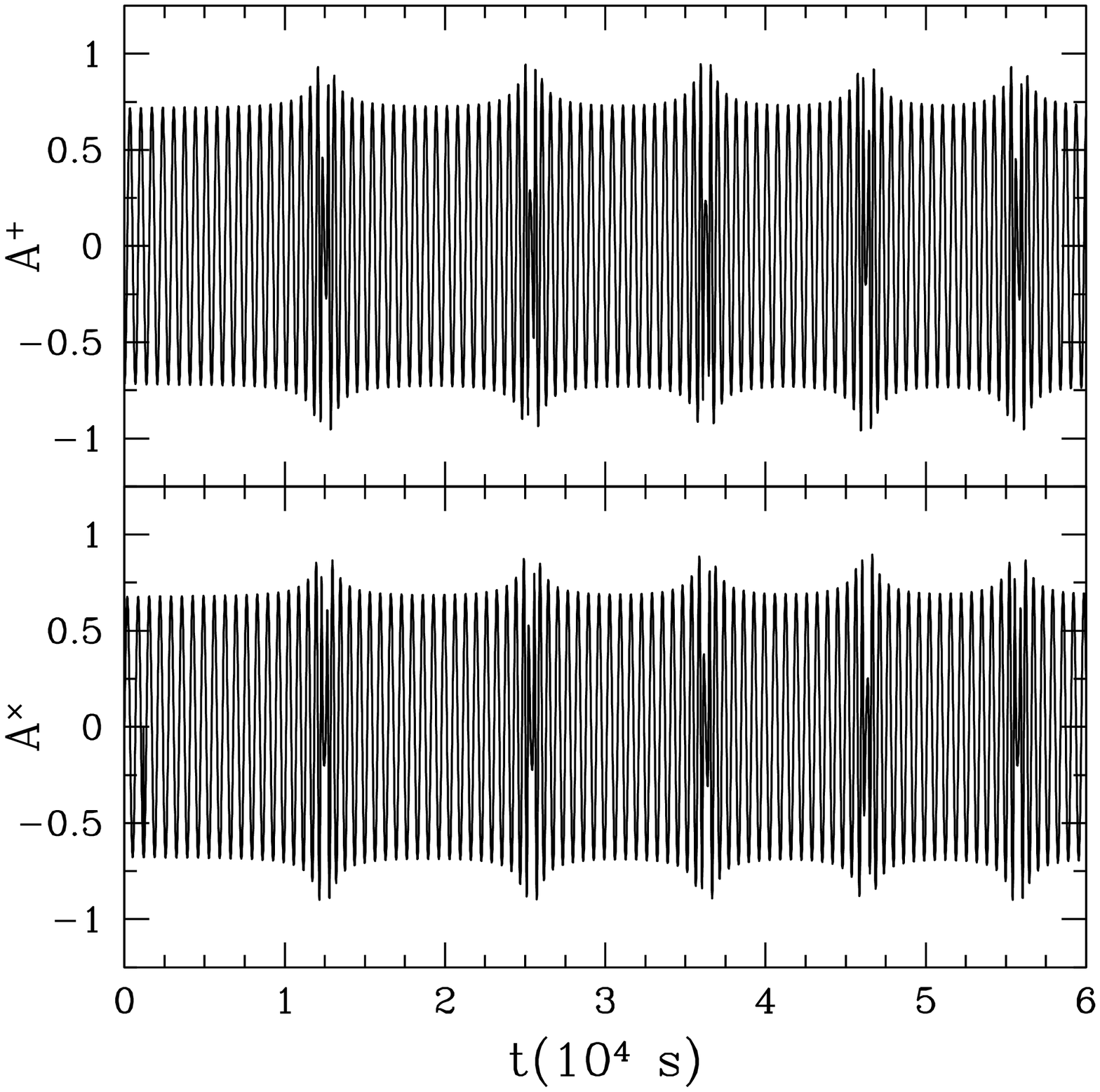}}
\scalebox{.50}{\includegraphics{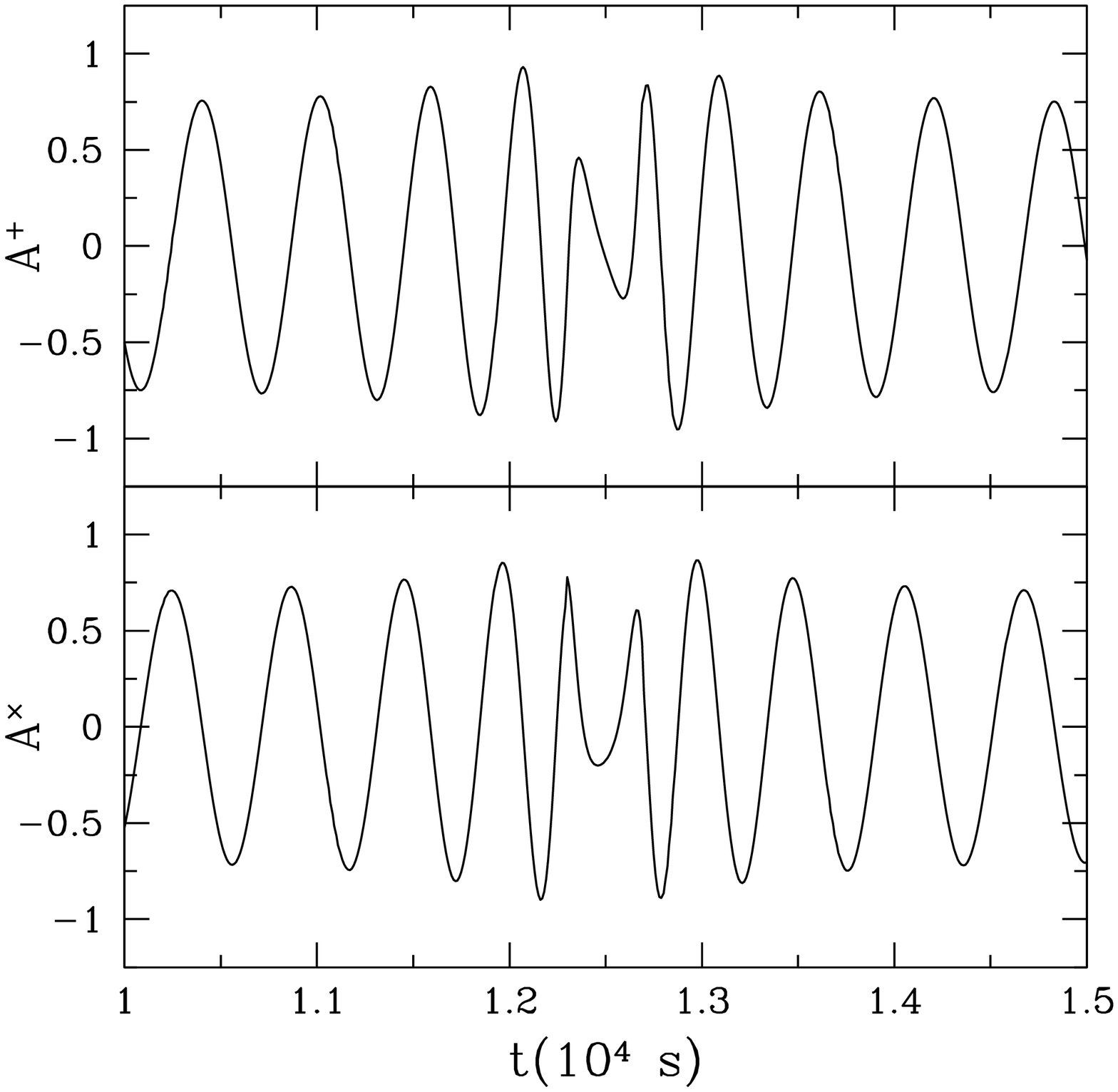}}
\caption{The GW amplitudes $\{ A^+(t), A^\times(t) \}$ for the initially
circular EMRI depicted in Fig.~\ref{F:xcirc}.  The dimensionless amplitudes are
given in units of $10^{-20}$, while the time is in units of $10^4$ s.  The
upper panel shows 60,000 s of the inspiral, including the plunge at
$t = 0.12 \times 10^4$ s.  The lower panel zooms in on the portion of the
waveform produced in the interior of the boson star itself.
\label{F:wcirc}}
\end{figure}

First we consider the waveform of the initially circular EMRI as shown in
Fig.~\ref{F:wcirc}.
The system is viewed at an inclination angle of $45^\circ$ at a distance of
10 Mpc.  The two polarizations provide similar information because the
given decomposition depends on our arbitrary choice of basis tensors.  Before
the plunge, the waveform is identical to that of a circular inspiral into a
Schwarzschild black hole.  The fundamental oscillation has a period of about
500 s, one-half the azimuthal period over which $\phi$ changes by $2\pi$
radians.  Since the orbit is circular, the amplitude only varies on the
extremely long timescale on which energy and angular momentum are lost.
After the plunge, the eccentricity increases sharply
and we see an amplitude modulation with a period of about 10,000 s.  This
corresponds to radial oscillations about the inner minimum within the boson
star.  The amplitude creeps upwards as the CO approaches pericenter with
increasing acceleration.  Note however the peculiar feature at pericenter
itself, where the amplitude should be greatest.  Instead, the cycle at
pericenter is suppressed because once the CO crosses the boson-star surface,
it is only accelerated by the mass interior to its position.  We present a
close-up of this distinctive feature in the bottom panel of Fig.~\ref{F:wcirc}.

\begin{figure}[t!]
\scalebox{.50}{\includegraphics{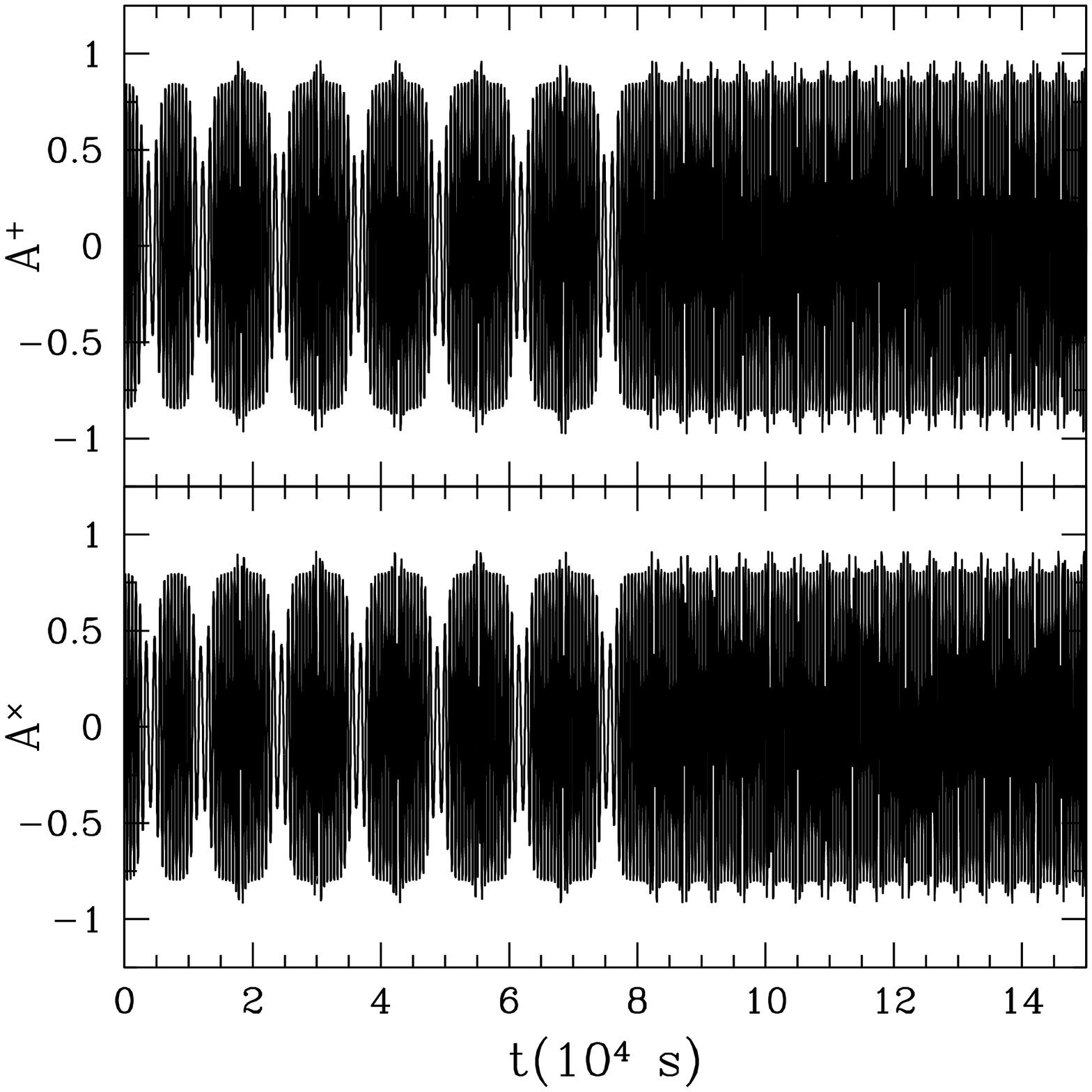}}
\caption{The GW amplitudes $\{ A^+(t), A^\times(t) \}$ for the initially
eccentric EMRI depicted in Fig.~\ref{F:xecc}.  As in Fig.~\ref{F:wcirc}, the
dimensionless amplitudes are given in units of $10^{-20}$, while the time is in
units of $10^4$ s.  All three portions of the inspiral are clearly visible.
\label{F:wecc}}
\end{figure}

\begin{figure}[t!]
\scalebox{.50}{\includegraphics{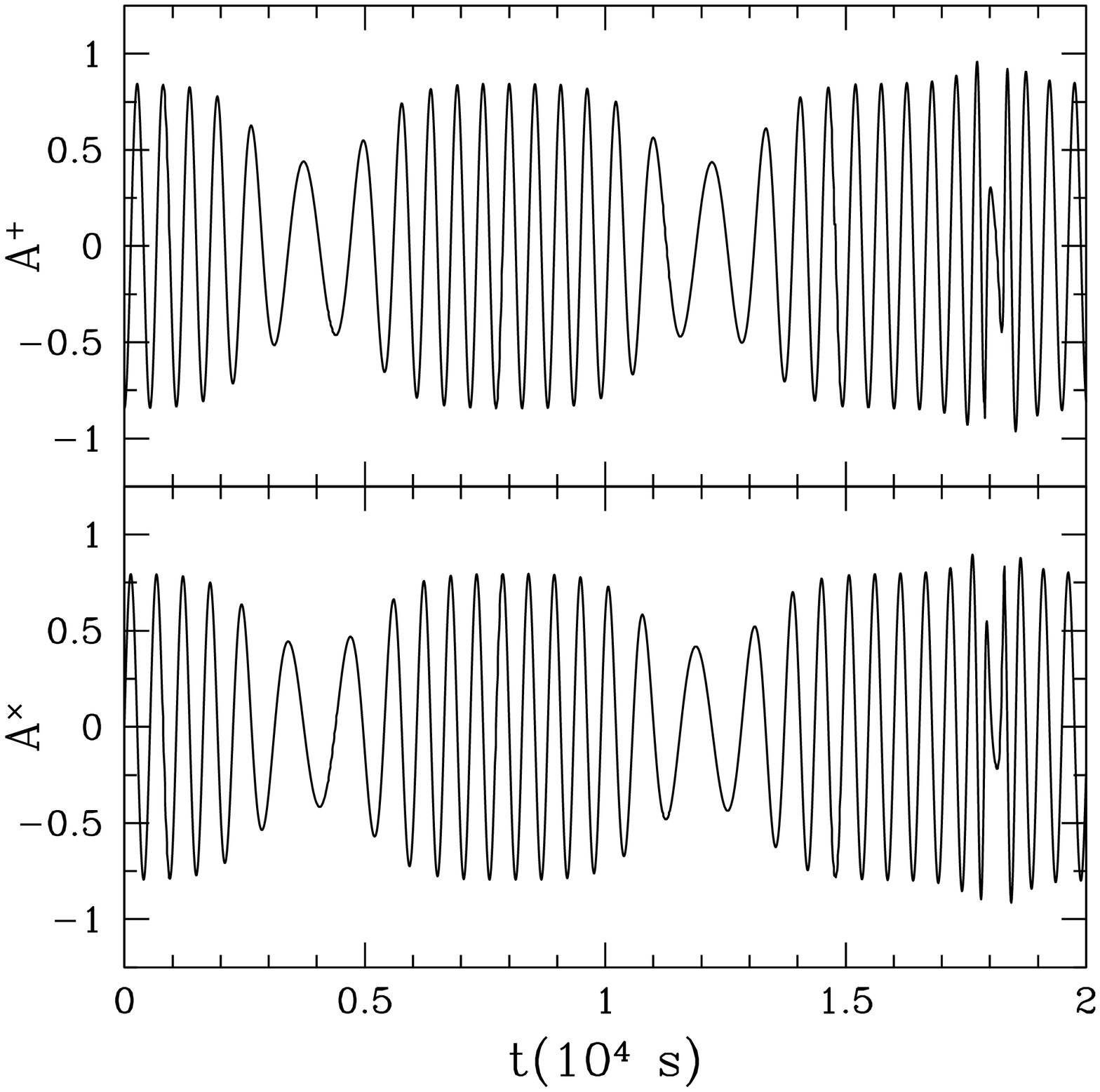}}
\scalebox{.50}{\includegraphics{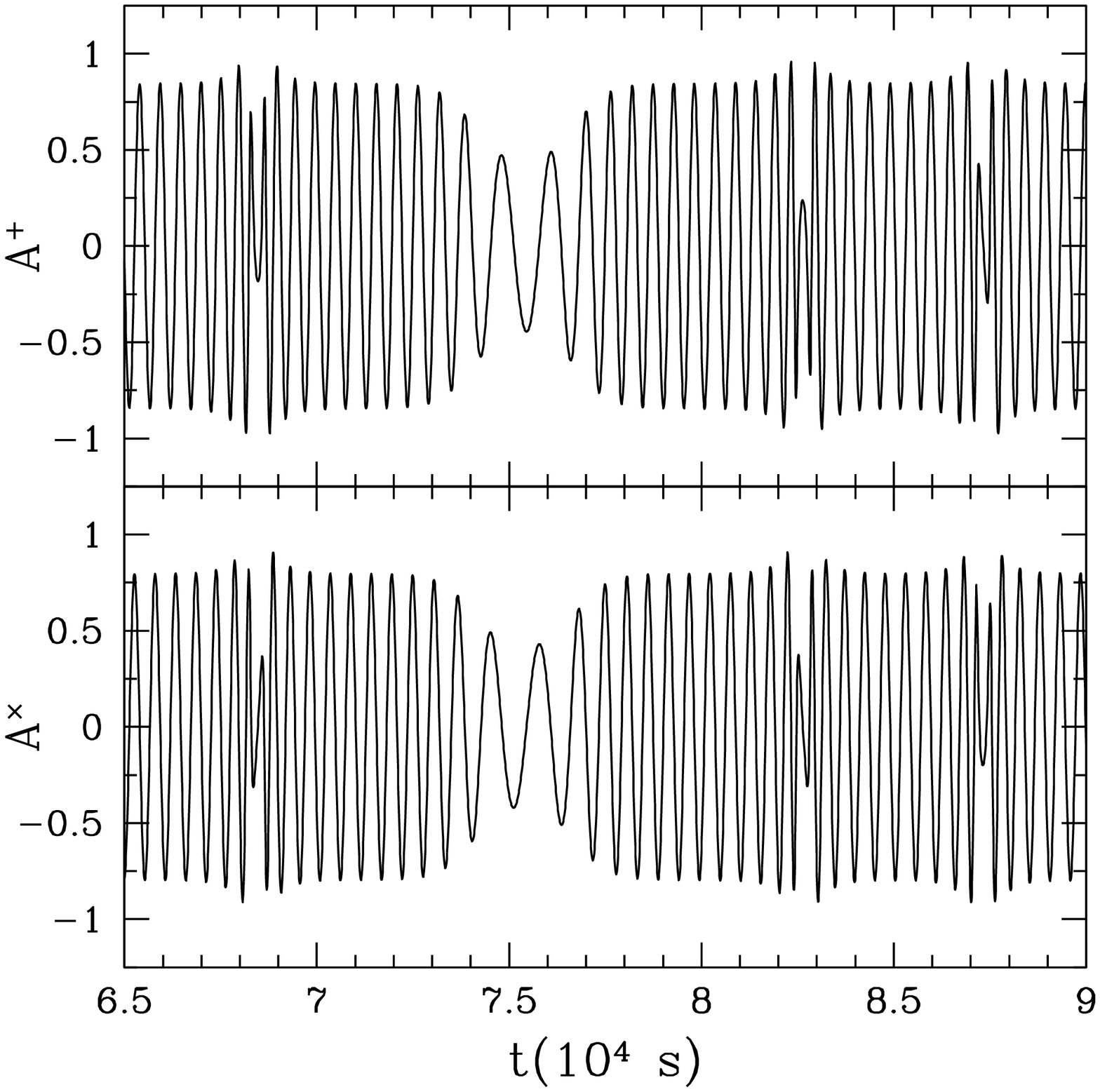}}
\caption{Close-ups of the waveform shown in Fig.~\ref{F:wecc} for the initially
eccentric EMRI.  The top panel depicts the first step of the plunge; note that
only the rightmost crest exhibits the peculiar feature produced in the
boson-star interior.  The bottom panel shows the second step of the plunge; note
how different the final distant apocenter at $t = 7.5 \times 10^4$ s appears from
the apocenter at $t = 8.5 \times 10^4$ s.
\label{F:TS}}
\end{figure}

The waveform of the initially eccentric EMRI as shown in Fig.~\ref{F:wecc} is
even more unique.  
All three phases of the EMRI can be easily distinguished by the naked eye.  In
the first phase, from the beginning of the waveform until the first-step of
the plunge at $t = 1.4 \times 10^4$ s, the CO moves on ordinary eccentric
geodesics
of Schwarzschild.  Amplitude modulation corresponding to oscillations about the
outer minimum are seen, the crests near pericenter are smooth because there are
as yet no close approaches to the boson star.  In the second phase, between the
two steps of the plunge, the CO moves on geodesics with {\it both} minima
between pericenter and apocenter.  These geodesics are extremely eccentric,
leading to longer radial oscillations and an approximate $50 \%$ increase in
the period of the amplitude modulation.  The apocenter remains nearly constant
during the plunge, marked by the continued preesnce of deep troughs in the
amplitude modulation.  The pericenter however has migrated deep into the
interior of the boson star as evidenced by the spikes on each crest produced
during close approaches.  The peculiar feature at pericenter itself is barely
visible on this scale as a very narrow gap in these spikes.  The second step
of the plunge occurs at $t=8.0\times10^4$ s, when the CO
crosses over the local maximum for
the final time.  The second step is in some sense the opposite of the first, 
in that the pericenter remains constant while the apocenter plunges inwards.
We see that the spikes marking pericenter remain unchanges between the second
and third phases, but the deep troughs in the amplitude modulation from distant
apocenters have vanished.  Close-ups of the two steps of the plunge are
shown in Fig.~\ref{F:TS}.

\section{Discussion} \label{S:disc}

Strong evidence supports the existence of supermassive compact objects at the
centers of many if not most galaxies.  Accretion onto these objects is presumed
to power active galactic nuclei, and direct observations of velocity dispersions
within galactic cusps reveal that a point-like mass dominates the dynamics
within $\sim 1$ pc from the center.  While all this evidence is consistent with
these objects being black holes, the ``smoking-gun'' signature of an event
horizon has yet to be observed.  Until such a definitive determination is made,
other candidates such as boson stars should continue to be considered.  Previous
work has suggested that accretion-induced X-ray bursts might be an optical
signature of boson stars \cite{XRB}.
We rely here on distinguishing boson stars from black
holes through their gravitational effects.  One possibility is their different
properties as a strong gravitational lens; a ``shadow'' induced by a black
hole's event horizon may soon be observed in Sag A$^\ast$ at submillimeter
wavelengths \cite{lense}.  In this paper, we have investigated
the possibility that
the inspiral of a several solar-mass CO into a boson star may produce a
distinctive spectrum of GWs.  LISA is conservatively expected to observe several
such inspirals each year with appreciable signal-to-noise ratios
\cite{Sig,SigRees}.

Previous work has examined the possibility of measuring the central object's
multipole moments, and testing whether or not they satisfy the predictions of
the ``no-hair'' theorem for black holes \cite{ryanMM}.  This formalism has been
applied in particular to boson stars with anomalously large mass-quadrupole
moments, where an appreciable effect on the GW-induced loss of energy was
discovered \cite{ryanBS}.  Here we have gone beyond such an
analysis by calculating approximations to the waveforms that might actually be
observed by LISA.
The spherically symmetric boson stars we consider have Schwarzschild spacetimes
outside their surfaces; a multipole analysis of GWs produced while the CO is still
outside the boson star would not reveal any violations of the ``no-hair'' theorem.
Solving
Einstein's equations exactly for the gravitational radiation produced would be
prohibitively expensive computationally.  Instead, we rely on a series of
analytic approximations: geodesic motion for the CO in the extreme mass-ratio
limit, GW propagation unaffected by the boson-star spacetime, and direct
indentification of Schwarzschild coordinates with flat-space spherical
coordinates.  These approximations, the ``particle on a string'' approach, allow
a direct application of multipole-moment expansions for both the energy and
angular momentum losses and the waveform itself.  We then drop all but the mass
quadrupole moment terms in these expansions.  These approximations
cannot be fully justified theoretically in the highly relativistic, strong-field
regime, but direct comparison with rigorous TSN calculations in the black hole
case shows much better agreement than one might have expected
\cite{GHK,TSNcomp}.  We then assume that the energy and angular momentum are
changing adiabatically on timescales much longer than an orbital period.  This
approximation, valid in the extreme mass-ratio limit except during the plunge
itself, allows us to apply a ``hybrid'' two-stage approach.  In the first stage,
differential equations for orbital elements $p$ and $e$ are obtained by relating
these quantities to $E$ and $L$ with exact analytic derivatives.  These
equations are then integrated on a radiation-reaction timescale, with a timestep
that is correspondingly long compared to an orbital
period.  In the second stage, the orbit of the CO, $r(t)$ and $\phi(t)$, is
computed with small timesteps by solving the geodesic equations and
linearly interpolating the trajectories $\{ p(t), e(t) \}$ produced in the
first stage.  The hybrid approach allows us to calculate real-space
trajectories and waveforms much more quickly than would be possible using the
same short timestep for both stages.

As anticipated, the waveforms produced by this method exhibit distinctive
features that allow them to be readily distinguished from those produced during
EMRIs into black holes.  In the model considered here, the boson-star inspiral
is identical to a black-hole inspiral until the CO falls over the
angular-momentum barrier. If LISA observed GWs from part of an inspiral
including this plunge, the parameters of the exterior black-hole spacetime
could be determined very accurately using black-hole EMRI templates from
the part of the inspiral up until plunge. The ``smoking gun'' for a boson-star
inspiral would be that GWs from the inspiral persist after the
plunge. This persistence could be seen using, for instance, a
time-frequency analysis of the LISA data stream. GWs from an event
like this could not be mistaken for an inspiral into a black hole with
different parameters, because the early stages of the inspiral are
identical to the black-hole inspiral. If only the post-plunge stage of a
boson-star inspiral were seen, it is not clear whether this could be mistaken
for a black-hole EMRI without a proper Fisher-matrix analysis.

The waveforms produced in this paper are highly approximate, far too crude to
use in any attempts at matched filtering for LISA.  Nonetheless, they may serve
as substitutes for purposes of scoping out the data analysis and design
specification until better waveforms are available.  Improved waveforms might
result from refining our method, for example, by incorporating higher-order
multipole moments in our expansions.  Press developed an improved formula for the
GW field $h_{ij}^{\rm TT}$ that accounts for time delays in the source, as well as
some of the relativistic effects provided by the higher moments \cite{press}.
Efforts are underway to adapt this formula to the ``particle on a string''
approach \cite{gairnotes,waveform}.  Analysis of higher harmonics of the
waveform might also prove to be a useful way of distinguishing EMRIs into black
holes and boson stars.  These higher harmonics become comparable to the
fundamental $m=2$ mode at high eccentricities \cite{BC1}.  One might expect the
sudden increase in eccentricity at plunge to manifest itself in frequency space
as a sharp rise in the higher harmonics.  Only a rigorous comparison of our
waveforms with the LISA noise curve can truly determine
whether LISA can differentiate between black-hole and boson-star EMRIs.
We hope to conduct such a comparison in the near future using
{\it Synthetic LISA}, an actual simulation of the LISA interferometer that models
its response to incident gravitational waves \cite{synLISA}.

Going beyond these approximations to produce waveforms suitable for
matched-filtering by LISA will involve the use of perturbation theory in the
the TSN formalism.  The resulting partial differential equations for a generic
boson star should be separable in the spherical case, and we plan to consider
this problem in the near future.  Hopefully, these
improvements or others will provide the tools to calculate accurate waveform
templates in time for LISA's great attempt at ``holiodesy'', the mapping
of spacetimes about the compact objects in galactic centers.  Including exotic
waveforms such as those produced by boson stars in our suite of templates will
ensure that LISA will not miss the opportunity to discover something truly
fundamental.

\begin{acknowledgments}
We wish to thank L. Lindblom and E. S. Phinney for useful conversations.  Micah
Solomon also provided help in preparing the Appendix.
Kesden was supported by the NASA Graduate Student Research Program.
JG's work was supported by NASA grants NAG5-12384 and NAG5-10707.  This work was
supported in part by DoE DE-FG03-92-ER40701 and NASA NAG5-9821.
\end{acknowledgments}

\appendix*
\section{Error Associated with the Quadrupole Approximation}
A key approximation made in \S~\ref{S:GR} was to drop all but the mass
quadrupole terms from the multipole-moment expansions of the GW field, energy
flux, and momentum flux in Eqs.~(\ref{E:hMM}), (\ref{E:EMM}), and (\ref{E:LMM}).
For circular orbits in the Schwarzschild metric, these expansions can be
expressed as power series in $M/r$ with coefficients that can be calculated
analytically.  To illustrate, the contribution of higher-order mass and current
multipole moments to the energy flux is given by
\begin{equation} \label{E:HO}
\frac{dE}{d(t/M)} = 
\begin{cases}
A_l \eta (r/M)^{-(l+3)}, &\text{mass multipoles} \\
B_l \eta (r/M)^{-(l+4)}, &\text{current multipoles}
\end{cases}
\end{equation}
where the dimensionless coefficients $A_l$ and $B_l$ \cite{mikemicah} are listed
in Table~\ref{T:coeff}.
\begin{table}
\caption{\label{T:coeff} Coefficients describing the contribution of higher-order
multipole moments to the energy flux of the inspiraling CO.}
\begin{ruledtabular}
\begin{tabular}{ccc}
$l$ & $A_l$ & $B_l$\\
\hline
2 & 6.400 & 1.067\\
3 & 6.458 & 4.571\\
2 & 14.966 & 1.250\\
\end{tabular}
\end{ruledtabular}
\end{table}
Neglecting the higher-order terms is clearly justified at large radii, but
the expansion parameter $M/r = 1/6$ is not particularly small at the ISO.  It is
still small enough however so that the next largest terms which happen to be the
$l=3$ and $l=4$ mass multipole moments provide only 16.8\% and 6.5\%
corrections respectively.  We continue to drop the higher-order terms as the CO
plunges past the ISO into the interior of the boson star.  If the singular
behavior of Eq.~(\ref{E:HO}) held within the boson star the expansion would
become formally divergent at $r = M$, but fortunately this is not the case.
Outside the boson star the angular frequency $\Omega$ is Keplerian,
$M\Omega = (r/M)^{-3/2}$, but in the interior $\Omega$ is a monotonically
increasing function of $r/M$.  As such, the higher-order terms
in the multipole-moment expansions which involve increasingly
more time derivatives become
increasingly steep functions of $r/M$.  The multipole-moment expansion is
therefore perfectly regular at the origin, and the quadrupole approximation
again becomes more accurate at small radii.  The errors associated with this
approximation, while excessive for the purposes of data analysis, are
acceptable for an initial search for qualitative features of a boson-star EMRI.

\bibliography{nov9} 

\end{document}